\documentclass[a4paper,12pt]{article}

\usepackage{epsfig}
\usepackage{amssymb,amsmath}
\newcommand{\tr}{\,\mbox{Tr}\,}
\newcommand{\lapprox}{%
\mathrel{%
\setbox0=\hbox{$<$}
\raise0.6ex\copy0\kern-\wd0
\lower0.65ex\hbox{$\sim$}
}}
\newcommand{\gapprox}{%
\mathrel{%
\setbox0=\hbox{$>$}
\raise0.6ex\copy0\kern-\wd0
\lower0.65ex\hbox{$\sim$}
}}

\newcommand{\ba}{\begin{array}}
\newcommand{\ea}{\end{array}}
\newcommand{\bd}{\begin{displaymath}}
\newcommand{\ed}{\end{displaymath}}
\newcommand{\be}{\begin{equation}}
\newcommand{\ee}{\end{equation}}
\newcommand{\bea}{\begin{eqnarray}}
\newcommand{\eea}{\end{eqnarray}}

\newcommand{\mm}{\mathcal{M}}
\newcommand{\diag}{\mbox{diag}\,}
\def\q2 {q^2}

\def\bt{\begin{table}}
\def\et{\end{table}}

\catcode`@=11 
\def \gsim{\mathrel{\mathpalette\@versim>}}
\def \lsim{\mathrel{\mathpalette\@versim<}}
\def \@versim#1#2{\lower0.4ex\vbox{\baselineskip\z@skip\lineskip\z@skip
     \lineskiplimit\z@\ialign{$\m@th#1\hfil##\hfil$%
     \crcr#2\crcr\sim\crcr}}}
\catcode`@=12 
\usepackage{rotating}
\usepackage{graphics}
\usepackage{subcaption}
\usepackage{float}
\restylefloat{table}
\begin{document}

\begin{flushright}
{\small 
RECAPP-HRI-2016-001} 
\end{flushright}

\begin{center}

{\large\bf A CP-violating phase in a two Higgs triplet scenario : some phenomenological implications}\\[15mm] 
Avinanda Chaudhuri \footnote{E-mail: avinanda@hri.res.in}
and Biswarup Mukhopadhyaya \footnote{E-mail: biswarup@hri.res.in} 
\\[2mm]
{\em Regional Centre for Accelerator-based Particle Physics \\
     Harish-Chandra Research Institute
\\
Chhatnag Road, Jhusi, Allahabad - 211 019, India}
 
\end{center}

\begin{abstract}
We consider a scenario where, along with the usual Higgs doublet, two scalar triplets are present. The extension of the triplet sector is required for the Type~II mechanism for the generation of neutrino masses, if this mechanism has to generate a neutrino mass matrix with two-zero texture. One CP-violating phase has been retained in the scalar potential of the model, and all parameters have been chosen consistently with the observed neutrino mass and mixing patterns. We find that a large phase ($\gtrsim 60^{\circ}$) splits the two doubly-charged scalar mass eigenstates wider apart, so that the decay $H_1^{++} \rightarrow H_2^{++} h$ is dominant (with h being the $125$ GeV scalar). We identify a set of benchmark points where this decay dominates. This is complementary to the situation, reported in our earlier work, where the heavier doubly-charged scalar decays as $H_1^{++} \rightarrow H_2^+ W^+$. We point out the rather spectacular signal, ensuing from $H_1^{++} \rightarrow H_2^{++} h$, in the form of Higgs plus same-sign dilepton peak, which can be observed at the Large Hadron Collider.

\end{abstract}

\vskip 1 true cm

\newpage
\setcounter{footnote}{0}

\def\baselinestretch{1.5}
\section{Introduction}

The observation of a rather distinctive pattern of neutrino mixing, together with
the available information on neutrino mass splitting, has triggered numerous
theoretical proposals going beyond the standard electroweak model (SM). 
Seesaw models enjoy a fair share of these, along with additional assumptions
to suit particular textures of the neutrino mass matrix. 

Type-II seesaw models can generate Majorana neutrino masses without
any right-handed neutrino(s), with the help of one or more Y=2 scalar triplets.
The restriction of the vacuum expectation value (vev) of such a triplet, arising from
the limits on the $\rho$-parameter (with $\rho = m^2_W/m^2_Z \cos^2 \theta$),
is obeyed in a not so unnatural manner, where the constitution of the scalar 
potential can accommodate large triplet scalar masses vis-a-vis a small vev.
In fact, this very feature earns such models classification as a type of `seesaw'.

A lot of work has been done on the phenomenology of scalar triplets which,
interestingly, also arise in left-right symmetric theories~\cite{goran}. One can,  however, still
ask the question: is a single-triplet scenario self-sufficient, or does the
replication of triplets (together with, say, the single scalar doublet of the SM)
bring about any difference in phenomenology? This question, otherwise a purely
academic one, acquires special meaning in the context of some neutrino mass
models which aim to connect the mass ordering with the values of the
mixing angles, thereby achieving some additional predictiveness.  One class of such models depend on texture zeros, where a number of zero entries (usually restricted to two) in the mass matrix enable one to establish the desired connection.
The existence of such zero entries require the imposition of some additional
symmetry; it has, for example, been shown that a horizontal ${\cal Z}_4$ symmetry
can serve the purpose. The simultaneous requirement of zero textures
and the type-II seesaw mechanism, however,  turns out to be inconsistent, as has
been discussed in earlier works~\cite{frampton}. The inconsistency is gone for two triplets.
This resurrects the relevance of the phenomenology of two-triplet, one doublet
scalar sectors, this time with practical implications.  We have studied such phenomenology in reference~\cite{avinanda}.  An important conclusion of this study was that,
whereas the doubly charged scalar in the single triplet scenario would decay mostly
in the $\ell^\pm \ell^\pm$ or $W^\pm W^\pm$ modes, the decay channel
$H^{++}_1 \longrightarrow H^{+}_2 W^+$ acquires primacy over a large region of
the parameter space. Some predictions on this in the context of the Large Hadron
Collider were also shown in reference~\cite{avinanda}. However, an added possibility with two
triplet is the possibility of at least one CP-violating phase being there.  This in principle can affect the phenomenology of the model, which is worth studying.

 With this in view, we have analyzed here the one-doublet, two-triplet framework,   including CP-violating effects  arising via a relative phase between the triplets. Thus, the vev of one triplet has been made complex and consequently, the coefficient of the corresponding trilinear term in the scalar potential has also been rendered complex.

 Indeed, the introduction of a phase results in some interesting findings that were not present when the relative phase was absent. First of all, as a result of mixing between two triplets and presence of a relative phase between them, something on which no phenomenological restrictions exits, the heavier doubly charged scalar can now decay frequently into the lighter doubly charged scalar plus the SM-like 
 Higgs boson, i.e 
 $H_1^{++} \rightarrow H_2^{++} h$, over a larger range of parameter space. This can give rise to a spectacular signal in the context of LHC. Basically, 
 as a final state, we obtain $H_1^{++} \rightarrow \ell^+ \ell^+ h$ i.e a doubly charged scalar decaying into two same-sign leptons plus the SM-like Higgs. This decay often dominates over all other decay channels. When this decay is not present due to insufficient mass-difference between respective scalars, the decay $H^{++}_2 \longrightarrow H^+_2 W^+$ mostly dominates, and its consequence was discussed in some detail in our earlier work~\cite{avinanda} on the CP-conserving scenario.
 
 Secondly, for some combination of parameters, the decays mentioned in the above paragraph are not possible with a vanishing or small phase, due to insufficient gap in masses between the respective scalars. However,  if we continuously increase the value of the  phase, keeping all the other parameters fixed, the mass difference between the scalars start in increase, so that the aforementioned channel  finally opens up.
  
 Thirdly, we have noticed in~\cite{avinanda} that the gauge coupling dominated decay $H^{++}_1 \rightarrow H^+_2 W^+$ dominates over the Yukawa coupling dominated decay $\Delta^{++} \rightarrow \ell^+ \ell^+$, even in those region of parameter space where we have chosen the Yukawa coupling matrices to be sufficiently large ($\simeq 1$). 
 On the other hand, the CP-violating phase suppresses the neutrino mass
 matrix elements for the same value of the triplet vacuum expectation values (vev). 
 This in turn requires an increase in the corresponding Yukawa coupling matrix elements, since the vev's and Yukawa couplings are related by the expression for
 neutrino masses. The outcome of this whole process is that,  for several BPs, the decay $H_1^{++} \rightarrow \ell^+ \ell^+$ competes with the decay into
 $H_2^+ W^+$.

 Finally, the CP-conserving scenario marks out regions of the parameter space, where the branching ratios of the decays 
 $H_1^{++} \longrightarrow \ell^+ \ell^+$ and $H_1^{++} \rightarrow W^+ W^+$  are of comparable, though subdominant,  rates. As the phase picks up, the same vev causes necessitates a hike in
Yukawa coupling, as discussed above. In such situations,  the decay 
$H_1^{++} \rightarrow \ell^+ \ell^+$ mostly dominates over the  $W^+ W^+$ mode. 

We present a summary of the scenario with a single triplet with complex vev
on section~2. In section~3, we outline the two-triplet scenario with a complex phase, including the corresponding scalar potential. The composition of physical states, the benchmark points for our numerical study and the results are presented in section~4. We summarize and conclude in section~5.

\section{A single scalar triplet with a CP-violating phase} 
\label{1-triplet}
We first give the reader a glimpse of the scenario
with a single triplet $\Delta = (\Delta^{++}, \Delta^{+}, \Delta^0)$, 
over and above the usual Higgs
doublet $\phi$, using the notation of~\cite{grimus3}.  
$\Delta$ is equivalently denoted by
the $2 \times 2$ matrix
\begin{equation}\label{triplet} 
\Delta = 
\left(\begin{array}{cc}
\Delta^+ & \sqrt{2}\Delta^{++} \\ \sqrt{2}\Delta^0 & -\Delta^+ 
\end{array}\right).
\end{equation}
The vev's of the doublet and the triplet
are expressed as
\begin{equation}\label{vev}
\langle \phi \rangle_0 =
\frac{1}{\sqrt{2}} \left(\begin{array}{c} 0 \\ v \end{array}\right)
\quad \mbox{and} \quad
\langle \Delta \rangle_0 =
\left(\begin{array}{cc} 0 & 0 \\ v_T & 0 \end{array}\right),
\end{equation}
respectively. The only doublet-dominated physical
state that survives after the generation of gauge boson masses is a 
neutral scalar $h$.

The most general scalar potential including $\phi$ and $\Delta$ can be
written as 
\begin{eqnarray}
V(\phi,\Delta)\  = \hphantom{+} a\, \phi^\dagger\phi + \ \frac{b}{2}
\tr(\Delta^\dagger\Delta)+ \ c\, (\phi^\dagger\phi)^2   
+ \ \frac{d}{4}\left( \tr(\Delta^\dagger\Delta) \right)^2 \nonumber \\ +
\ \frac{e - h}{2} \phi^\dagger\phi \tr(\Delta^\dagger\Delta)  
+  \ \frac{f}{4} \tr(\Delta^\dagger\Delta^\dagger)\tr(\Delta\Delta)
\nonumber \\ 
+ \ h \phi^\dagger \Delta^\dagger \Delta \phi + \left( t\, \phi^\dagger
\Delta \tilde{\phi} + 
\mbox{H.c.} \right),
\end{eqnarray}
where $\tilde{\phi} \equiv i\tau_2 \phi^\ast$.
All parameters in the Higgs potential are real except $t$ which is complex in general. By performing a global $U(1)$ transformation, $v$ can always be chosen real and positive. Because of the $t$-term in the potential there is no second global symmetry  to make $v_T$ real. Furthermore, $t$ can also be complex and, therefore, it can be written as $t = \lvert t\rvert\ e^{i\alpha}$ and $v_T = w e^{i\gamma}$ with $w \equiv \lvert v_T\rvert $. Minimization of the scalar potential with
respect to the phase of $v_T$ i.e $\gamma$, gives the relation between the phases as $\alpha + \gamma = \pi$. \footnote{For an analogous situation with two Higgs doublets, see, for example~\cite{ma}.}

The choice $a<0$, $b>0$ ensures that the dominant source of spontaneous
symmetry breaking is the scalar doublet. It is further assumed,
following~\cite{grimus3}, that
\begin{equation}\label{oomagn}
a,\: b \sim v^2; \quad
c,\: d,\: e,\: f,\: h \sim 1; \quad 
|t|\ll v.  
\end{equation}
Such a choice is motivated by the following considerations
\begin{itemize}
\renewcommand{\labelenumi}{(\alph{enumi})}
\item
The need to fulfill  the
electroweak symmetry breaking conditions, 
\item
To have $w \ll v$ sufficiently small, as required by
the $\rho$-parameter constraint, 
\item
To keep doublet-triplet mixing low in general,
and 
\item 
To ensure that all quartic couplings are perturbative.  
\end{itemize}

The mass terms for the singly-charged scalars can be expressed in a compact form as
\begin{equation}
\mathcal{L}^\pm_S = -
(H^-, \phi^-) \mathcal{M}_+^2 
\left( \begin{array}{c} H^+ \\ \phi^+ \end{array} \right)
\end{equation}
with
\begin{equation}
\mathcal{M}_+^2 = \left(\begin{array}{cc}
(q+h/2)v^2 & \sqrt{2} v (t^* - v_T h/2) \\
\sqrt{2} v (t - v_T^* h/2) & 2(q + h/2)w^2 
\end{array}\right) 
\quad \mbox{and} \quad 
q = \frac{|t|}{w}.
\end{equation}
Keeping aside the charged Goldstone boson, the mass-squared of
the singly-charged physical scalar is obtained as
\begin{equation}
m^2_{\Delta^+} = \left(q + \frac{h}{2} \right) (v^2 + 2w^2),
\end{equation} 
while the doubly-charged scalar mass is expressed as
\begin{equation}
m^2_{\Delta^{++}} = (h + q) v^2 + 2fw^2.
\end{equation}
Thus, in the limit  $w \ll v$,
\begin{equation}
m^2_{\Delta^{++}} - m^2_{\Delta^+} \simeq {\frac {h}{2}}v^2.
\end{equation}

Thus a substantial mass splitting 
between $\Delta^{++}$ and $\Delta^+$ is in general difficult.
This tends to disfavour the $\Delta^+ W^+$ decay channel of
$\Delta^{++}$, as compared to $\ell^+ \ell^+$ and $W^+ W^+$.

\section{Two scalar triplets and a CP-violating phase}
\label{2-triplet}
Strong evidence has accumulated in favour of neutrino oscillation
from the solar, atmospheric, reactor and accelerator neutrino
experiments over the last few years. It is now widely believed that
neutrinos have non-degenerate masses and a very characteristic
mixing pattern. A lot, however,  is yet to be known, including the mass generation mechanism
and the absolute values of the masses, as opposed to mass-squared differences
which affect oscillation rates. Also, a lot of effort is on
to ascertain the nature of neutrino mass hierarchy, including the
signs of the mass-squared differences. A gateway to information of the
above kinds is the light  neutrino  mass matrix, in a basis
where the charged lepton mass matrix is diagonal.

 Here, too, in the absence of very clear guidelines, various `textures' for
the neutrino mass matrix are often investigated. A possibility that
frequently enters into such investigations is one where the mass matrix has
some zero entries, perhaps as the consequence of some built-in symmetry 
of lepton flavours.  At the same time, such `zero textures'  lead to 
a higher degree of predictiveness and inter-relation between mass
eigenvalues and mixing angles, by virtue of having fewer free paramaters, see for example~\cite{zero}.
In the context of Majorana neutrinos which have a symmetric
mass matrix, various texture zeros have thus been studied 
from a number of angles. Of them, two-zero textures have a rather
wide acceptability.

 It has been shown in~\cite{grimus4} that none of the seven possible
two-zero-texture cases can be achieved by assuming only one scalar triplet.
However, this is not the case with two triplets,  where several 
of the seven possible two-zero textures are allowed. Therefore, it is
important to examine the phenomenological consequences of an
augmented triplet sector, if Type-II seesaw has to be consistent
with the texture-zero approach.

'
One is thus encouraged to consider a scenario consisting of  
one complex doublet and two 
$Y=2$ triplet scalars $\Delta_1$, $\Delta_2$, both written as 
2$\times$2 matrices:
\begin{equation}\label{triplet2}
\Delta_1 = 
\left(\begin{array}{cc}
\delta_1^+ & \sqrt{2}\delta_1^{++} \\ \sqrt{2}\delta_1^0 & -\delta_1^+ 
\end{array}\right)
\quad \mbox{and} \quad
\Delta_2 = 
\left(\begin{array}{cc}
\delta_2^+ & \sqrt{2}\delta_2^{++} \\ \sqrt{2}\delta_2^0 & -\delta_2^+ 
\end{array}\right).
\end{equation}
The vev's of the scalar triplets are given by
\begin{equation}\label{vev2'}
\langle \Delta_1 \rangle_0 =
\left(\begin{array}{cc} 0 & 0 \\ w_1 & 0 \end{array}\right) 
\quad \mbox{and} \quad
\langle \Delta_2 \rangle_0 =
\left(\begin{array}{cc} 0 & 0 \\ w_2 & 0 \end{array}\right).
\end{equation}
The vev of the Higgs doublet is as usual given by equation~(\ref{vev}).

The scalar potential in this model involving $\phi$, $\Delta_1$
and $\Delta_2$ can be written as 
\begin{eqnarray}
\lefteqn{V(\phi,\Delta_1,\Delta_2) =}
\nonumber \\
&&
a\, \phi^\dagger\phi + 
\frac{1}{2}\, b_{kl} \tr(\Delta_k^\dagger \Delta_l)+ 
c (\phi^\dagger\phi)^2 + 
\frac{1}{4}\, d_{kl} \left( \tr(\Delta_k^\dagger\Delta_l) \right)^2
\nonumber \\
&&
+ \frac{1}{2}\,(e_{kl} - h_{kl})\,
\phi^\dagger \phi  \tr (\Delta_k^\dagger\Delta_l) +
\frac{1}{4}\,f_{kl}  
\tr(\Delta_k^\dagger\Delta_l^\dagger) \tr(\Delta_k\Delta_l)
\nonumber \\
&&
+ h_{kl}\, \phi^\dagger \Delta_k^\dagger \Delta_l \phi +
g \tr(\Delta_1^\dagger\Delta_2) \tr(\Delta_2^\dagger\Delta_1) + 
g' \tr(\Delta_1^\dagger\Delta_1) \tr(\Delta_2^\dagger\Delta_2) 
\nonumber \\
&&
+ \left( t_k\, \phi^\dagger \Delta_k \tilde{\phi} +
\mbox{H.c.} \right),
\label{pot}
\end{eqnarray}
where summation over $k,l=1,2$ is understood.
This potential is not the most general one, since we neglected some of
the quartic terms. This is justified in view of the scope of this
paper, as laid out in the introduction. 

 In~\cite{avinanda},  all the vev's as well the parameters in the potential
 were assumed to be real. As has already ben mentioned, this need not
 be the situation in general. To see the phenomenology including CP-violation,
 we make a minimal extension of the simplified scenario by postulating {\em one}
CP-violating phase to exist. This entails a complex vev for any one triplet 
(in our case we have chosen it to be $\Delta_1$). 
At the same time, there is a complex phase in the coefficient $t_1$ of the trilinear 
term in the potential.  Thus one can write $t_1 = \lvert t_1\rvert\ e^{i\beta}$ and 
$w_1 = \lvert w_1\rvert\ e^{i\alpha}$.  

Using considerations very similar to those for the single-triplet model,
we have taken
\begin{equation}\label{oomagn2}
a,\: b_{kl} \sim v^2; \quad
c,\: d_{kl},\: e_{kl},\: h_{kl},\: f_{kl},\: g,\: g' \sim 1; \quad
|t_k| \ll v. 
\end{equation}
We also chosen to restric ourselves to cases where  $w_1, w_2 \ll v$, keeping in mind
the constraint on the $\rho$-parameter.

 The mass eigenvalues, scalar mixing matrices etc.following from the
 potential~(\ref{pot}) can only be obtained numerically in general.
However, one can use the smallness of the triplet vev's
$w_k$, and  drop the quartic terms in the scalar triplets during
the diagonalisation of the mass matrices.
This enables one to use approximate analytical expressions, which
makes our broad conclusions somewhat transparent. However, 
the numerical results 
presented in section~\ref{results1} are obtained using the full
potential~(\ref{pot}), including the effects of the  triplet vev's. 

It is convenient to speak in terms of the following matrices and vectors: 

\begin{eqnarray}
B = (b_{kl}), \quad \quad E = (e_{kl}), \quad \quad H = (h_{kl}),  
\end{eqnarray} 
\begin{eqnarray}
t = \left( \begin{array}{c} \lvert t_1\rvert cos\beta \\ t_2 \end{array} \right),
t' = \left( \begin{array}{c} \lvert t_1\rvert sin\beta \\ 0 \end{array} \right),
w = \left( \begin{array}{c} \lvert w_1\rvert cos\alpha \\ w_2 \end{array} \right),
w' = \left( \begin{array}{c}\lvert w_1\rvert sin\alpha \\ 0 \end{array} \right).
\end{eqnarray}

In terms of  them, the conditions for a stationary point of the
potential are 
\begin{eqnarray}
\left( B + \frac{v^2}{2} \left( E-H \right) \right) w + 
v^2\, t &=& 0, 
\label{vev1} \\
a + cv^2 + \frac{1}{2} w^T (E-H) w + 2\, t \cdot w + 2 t' \cdot w' + \frac{1}{2} w'^T (E-H)w' &=& 0,
\label{vev2} \\
(b_{11} + \frac{v^2}{2} (e_{11}-h_{11})) \lvert w_1\rvert sin\alpha - v^2 \lvert t_1\rvert sin\beta &=& 0,
\label{vev3}
\end{eqnarray}
using the notation $t \cdot w = \sum_k t_k w_k$. These three equations are exact if one neglects all terms quartic in the
triplet vev's in $V_0 \equiv V( \langle \phi \rangle_0, \langle \Delta \rangle_0)$.
In equation~(\ref{vev2}) we have already divided by $v$, assuming 
$v \neq 0$. The small vev's $w_k$ are thereafter obtained as
\begin{equation}
w = -v^2 \left( B + \frac{1}{2} v^2 (E-H) \right)^{-1} t.
\end{equation}
And from equation~(\ref{vev3}) the phase of $t_1$ can be expressed  as
\begin{equation}\label{phase1}
sin\beta = \frac{v^{-2}(b_{11} + \frac{v^2}{2} (e_{11}-h_{11})) \lvert w_1\rvert sin\alpha}{\lvert t_1\rvert}
\end{equation}
This re-iterates the fact that the phases  $t_1$ and  $w_1$ are  related to each other. It is also evident from~(\ref{phase1}) that the value of the angle $\alpha$ has to be $n\pi$ where $n = 0, 1, 2, 3 ....$ when the phase  $\beta$ is absent.

 We next discuss the mass matrices of charged scalars. The mass matrix of the doubly-charged scalars is obtained as
\begin{equation}
\mm^2_{++} = B + \frac{v^2}{2} \left( E+H \right).
\end{equation}
It is interesting to note that if we drop those quartic terms for simplification from our scalar potential, then our doubly charged mass matrix remains the same as in~\cite{avinanda}. This gives the impression that the relative phase between triplets does not affect the doubly charged mass matrix if we drop the quartic terms in the potential. But in our numerical calculation, where we have taken the full scalar potential including the quartic terms, we find such a dependence, arising
obviously from the quartic terms.  This will be discussed further in the next section.

As for the singly-charged fields $\Delta^+_k$, one has consider their mixing 
with $\phi^+$ of the Higgs doublet. and This  introduces the CP-violating phase   
into the singly charged mass matrix. We write
the mass term as 
\begin{equation}\label{m++}
-\mathcal{L}^{\pm}_S = 
\left( \delta^-_1, \delta^-_2, \phi^- \right) \mm^2_+
\left( \begin{array}{c}
\delta^+_1 \\ \delta^+_2 \\ \phi^+ \end{array} \right)
+ \mbox{H.c.},
\end{equation}
equation~(\ref{pot}) leads to 
\begin{equation}\label{m+}
\renewcommand{\arraystretch}{1.2}
\mm^2_+ = \left(
\begin{array}{cc}
B + \frac{v^2}{2} \, E & 
\sqrt{2} v \left( t - H w/2 \right) \\
\sqrt{2} v \left( t - H w/2 \right)^\dagger &
a + cv^2 + \frac{1}{2} w^T (E+H) w + \frac{1}{2} w'^T (E+H) w'
\end{array} \right).
\end{equation}
Now, this mass matrix must have a zero eigenvalue,
corresponding to the would-be-Goldstone boson. Indeed, 
on substituting the minimization equations~{(\ref{vev1}), (\ref{vev2})} and~(\ref{vev3}), we see that
\begin{equation}\label{wouldbe}
Det(\mm^2_+) = 0,
\end{equation}
which ensures a consistency check.

The mass matrices~(\ref{m++}) and~(\ref{m+}) are diagonalized by
\begin{equation}\label{UV}
U_1^\dagger \mm^2_{++} U_1 = \diag (M^2_1, M^2_2) 
\quad \mbox{and} \quad
U_2^\dagger \mm^2_{+} U_2 = \diag (\mu_1^2, \mu_2^2, 0),
\end{equation}
respectively, with
\begin{equation}\label{UV1}
\left( \begin{array}{c} \delta^{++}_1 \\ \delta^{++}_2
\end{array} \right) = U_1
\left( \begin{array}{c} H^{++}_1 \\ H^{++}_2
\end{array} \right), 
\quad
\left( \begin{array}{c} \delta^{+}_1 \\ \delta^{+}_2 \\ \phi^+
\end{array} \right) = U_2
\left( \begin{array}{c} H^{+}_1 \\ H^{+}_2 \\ G^+
\end{array} \right).
\end{equation}
We have denoted the fields with definite mass by $H^{++}_k$ and
$H^{+}_k$, and $G^+$ is the charged would-be-Goldstone boson.

We also outline the neutral sector of the model, which cannot now be separated into CP-even and CP-odd sectors. Thus the mass matrix for the neutral sector of the present scenario turns out to be a $6 \times 6$ matrix , including mixing between real and imaginary parts of the complex neutral fields. The symmetric neutral mass matrix is denoted by $\mathcal{M}_{neut}$, whose elements are listed in the Appendix. So, the mass term for the neutral part can be written as :
\begin{equation}\label{mneut}
-\mathcal{L}^0_S = 
\left( N_{01}, N_{02}, N_{03}, N_{04}, N_{05}, N_{06} \right) \mathcal{M}^2_{neut}
\left( \begin{array}{c}
N_{01} \\ N_{02} \\ N_{03} \\ N_{04} \\ N_{05} \\ N_{06} \end{array} \right)
+ \mbox{H.c.},
\end{equation}
Where $N_{0n}$ s are the neutral states in flavor basis.  This mass matrix is diagonalized by
\begin{equation}\label{mneut1}
U_3^\dagger \mathcal{M}^2_{neut} U_3 = \diag (M^2_{01}, M^2_{02}, M^2_{03}, M^2_{04}, M^2_h, 0) 
\end{equation}
with
\begin{equation}\label{UV1}
\left( \begin{array}{c} N_{01} \\ N_{02} \\ N_{03} \\ N_{04} \\ N_{05} \\ N_{06}
\end{array} \right) = U_3
\left( \begin{array}{c} H_{01} \\ H_{02} \\ H_{03} \\ H_{04} \\ h \\ G_0
\end{array} \right). 
\end{equation}
Where $h$ is identified with the Standard Model Higgs boson and $G_0$ is the neutral Goldstone boson.

 It is interesting to note that once we remove the phases of coefficient of trilinear term in scalar potential and the vev of triplet $H_1^{++}$ 
by setting  $\alpha = \beta = 0$ , then the mixing between the CP-even and CP-odd scalars vanishes and we get back the usual separate $3 \times 3$ matrices for these two sectors. This also serves as a consistency check for the model. And ofcourse the lightest neutral scalar of this sector can be identified with SM Higgs.

 The $\Delta L = 2$ Yukawa interactions of the triplets are  
\begin{equation}\label{Ll}
\mathcal{L}_Y = \frac{1}{2} \,
\sum_{k=1}^2 y^{(k)}_{ij} L^T_i C^{-1} i \tau_2 \Delta_k L_j + 
\mbox{H.c.},
\end{equation}
where $C$ is the charge conjugation matrix, the $y^{(k)}_{ij}$ 
are the symmetric Yukawa coupling matrices of the triplets
$\Delta_k$, and the $i,j$ are the
summation indices over the three neutrino flavours.
  The charged-lepton mass matrix is diagonal in this basis.

 The neutrino mass matrix is generated from $\mathcal{L}_Y$ as
\begin{equation} \label{numas}
(M_\nu)_{ij}
= y^{(1)}_{ij} \lvert w_1\rvert cos\alpha + y^{(2)}_{ij} w_2.
\end{equation}
This relates the Yukawa coupling constants $y^{(1)}_{ij}$, 
$y^{(2)}_{ij}$ and the real part of the triplet vev's, namely, 
$\lvert w_1\rvert cos\alpha$ and $w_2$.

The neutrino mass eigenvalues 
are fixed according to a particular type of mass spectrum.
In this work we illustrate our points, without any loss 
of generality, in the context of normal hierarchy,  setting the lowest
neutrino mass eigenvalue to zero.  
Next, using the observed central values of the various 
lepton mixing angles, 
the elements of the neutrino mass matrix $M_\nu$ can be found by using
\begin{equation}\label{Mnu}
M_\nu = U^\dagger {\hat{M}}_\nu U,
\end{equation} 
where $U$ is the PMNS matrix given by~\cite{pdg} 
\begin{equation}
U = 
\left(\begin{array}{ccc}
c_{12} c_{13} & s_{12} c_{13} & s_{13} e^{-i\delta}  \\ - s_{12} c_{23} -
c_{12} s_{23} s_{13} e^{i\delta}  & c_{12} c_{23} - s_{12} s_{23} s_{13}
e^{i\delta} & s_{23} c_{13}  \\ s_{12} s_{23} - c_{12} c_{23} s_{13}
e^{i\delta} & - c_{12} s_{23} - s_{12} c_{23} s_{13} e^{i\delta} & c_{23}
c_{13} 
\end{array}\right)
\end{equation}
and ${\hat{M}}_\nu$ is the diagonal matrix of the neutrino masses.
We have neglected  possible Majorana phases, and
the recent global analysis of neutrino data are used to compute the
elements of $U$~\cite{fogli}.  Also, the phase  $\delta$ has been set
to  zero.  For $\theta_{13}$,  the results from the 
Daya Bay and RENO experiments~\cite{dayabay,RENO} have been used.

After all this, all terms of the left-hand side of equation~(\ref{Ll}) 
are approximately known, which is sufficient for predicting
phenomenology in the 100 GeV - 1 TeV scale..
The actual mass matrix thus constructed, on  numerical evaluation,
approximately reflects a two-zero texture which is one of the motivations
of this study. 

For each  benchmark point used in the next section, $w_1$ and
$w_2$ get determined by values of the other
parameters in the scalar potential.  Of course, the coupling matrices
$y^{(1)}$ and $y^{(2)}$ are still indeterminate. 
We fix the  matrix $y^{(2)}$
by choosing a single suitable value for all elements of the 
$\mu$--$\tau$ block and a smaller value 
for the rest of the matrix. 
As has already been
mentioned in~\cite{avinanda},  our broad conclusions do not depend on this `working rule'.

%
\section{Benchmark points and numerical predictions}
\label{results1}
The trademark signal of Higgs triplets is contained in the doubly charged 
components. In the current scenario, too, one would like to see the signatures
of the two doubly charged scalars, especially the heavier one, namely  $H^{++}_1$ 
whose decays have already been shown to contain a rather rich phenomenology.

The $H^{++}_1$, produced at the LHC via the Drell-Yan process, can in general have two-body decays in the following channels:

\begin{eqnarray}
H^{++}_1 &\rightarrow& H^{++}_2 h,  \label{h50} \\
H^{++}_1 &\rightarrow& \ell^+_i \ell^+_j, \label{hll} \\
H^{++}_1 &\rightarrow& W^+ W^+, \label{hww}   \\
H^{++}_1 &\rightarrow& H^+_2  W^+,  \label{hhw} \\
H^{++}_2 &\rightarrow& \ell^+_i \ell^+_j, \label{h2ll} \\
H^{++}_2 &\rightarrow& W^+ W^+, \label{h2ww}
\end{eqnarray}
with  $h$ is the SM-like Higgs  and $\ell_i, \ell_j = e, \mu$.

 The decay modes (\ref{h50})and~(\ref{hhw}) are absent in the
 single-triplet model. On the other hand, mixing between two triplets opens up situations where the mass separation between $H^{++}_1, H^{++}_2$ and $H^{++}_1, H^{+}_2$ is sufficient to kinematically allow the transitions~(\ref{h50}) and~(\ref{hhw}). The decay~(\ref{h50}) opens up a spectacular signal, especially when $H^{++}_2$ mostly decays into two same sign leptons, leading to
\begin{equation} \label{hll2}
H^{++}_1 \rightarrow \ell^+_i \ell^+_j h
\end{equation}
Let us denote the mass of SM Higgs by $M_h$, that of $H_k^{++}$ by $M_k$ and that of $H_k^+$ by $\mu_k$ ($k=1,2$). 
Then, in the convention $M_1 > M_2$,  $\mu_1 > \mu_2$, 
the decays~(\ref{h50})and~(\ref{hhw}) are possible only if $M_1 > M_2 + M_h$ and $M_1 > \mu_2 + m_W$.
We demonstrate numerically that this can naturally happen, by 
considering three distinct regions of the parameter space and selecting
four benchmark points (BPs) for each region. The relative phase between two triplets also a plays an important roll in these cases. In order to
emphasise this, we have also chosen three different values of the phase, namely $\alpha = 30^{\circ}, 45^{\circ}$ and $60^{\circ}$ for each  benchmark point. Thus we have considered 36 BPs alltogether, comprising three distinct regions of the parameter space and relative phases between triplets to justify our findings.

 We have seen that, in a single-triplet model, the
doubly-charged Higgs decays into either $\ell^+_i \ell^+_j$ or $W^+
W^+$. The former is controlled by the $\Delta L =2$ Yukawa couplings
$y_{ij}$, while the latter is driven by $w$, the triplet vev. 
Neutrino masses are given by~(\ref{numas}), implying large
values of $y_{ij}$ for small $w$, and vice versa. Interestingly, 
the presence of triplet phase through the $cos\alpha$ term in this equation actually suppresses the vev $w_1$ of the first triplet. This in turn implies that we get higher values for Yukawa coupling matrix entries $y_1^{ij}$ compared to the case  where CP-violating effects are  absent. 
Accordingly, we have identified, for the chosen values of triplet phase, three regions in the parameter space, corresponding to
\begin{enumerate}
\renewcommand{\labelenumi}{\roman{enumi})}
\item
$\Gamma(H^{++}_{1,2} \rightarrow \ell^+_i \ell^+_j ) \ll
\Gamma(H^{++}_{1,2} \rightarrow W^+ W^+ )$, 
\item
$\Gamma(H^{++}_{1,2} \rightarrow \ell^+_i \ell^+_j ) \gg
\Gamma(H^{++}_{1,2} \rightarrow W^+ W^+ )$, 
\item
$\Gamma(H^{++}_{1,2} \rightarrow \ell^+_i \ell^+_j ) \sim
\Gamma(H^{++}_{1,2} \rightarrow W^+ W^+ )$. 
\end{enumerate}
These are referred to as scenarios 1, 2 and 3 respectively in the
subsequent discussion.

The masses of the various physical state scalars 
are shown in Tables. Although our study 
involves mainly  the phenomenology of charged scalars, we have also listed the masses of neutral scalars. It should be noted that the lightest neutral scalar, dominated by the doublet component, has mass $\sim 125$ GeV for each BP, identifying it with the observed Higgs particle.

\begin{figure}
\centering

\begin{subfigure}[t]{.4\textwidth}
\centering
\includegraphics[angle=-90,width=\linewidth]{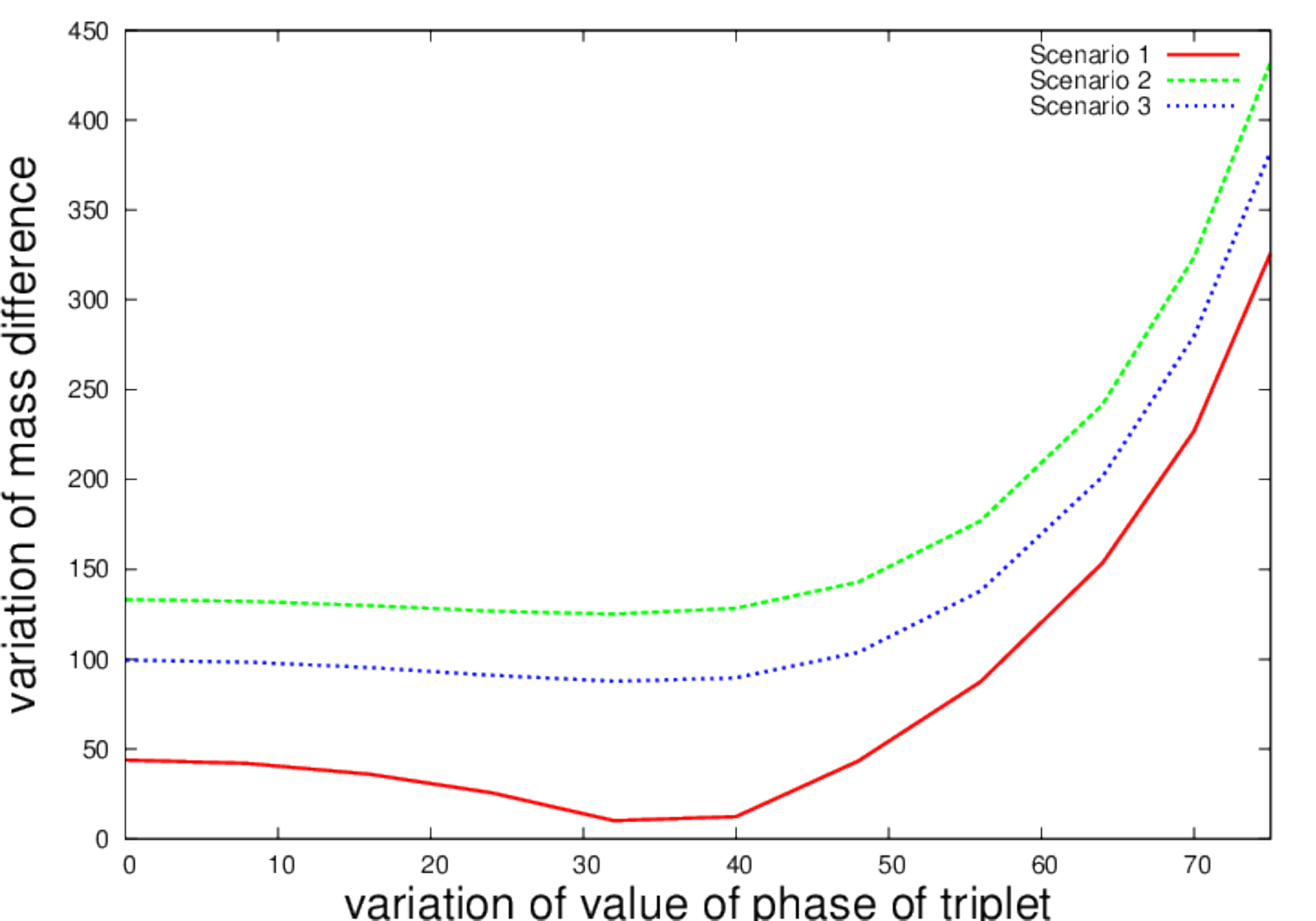}
        \caption{}\label{fig:fig_f}
\end{subfigure}
\begin{subfigure}[t]{.4\textwidth}
\centering
\includegraphics[angle=-90,width=\linewidth]{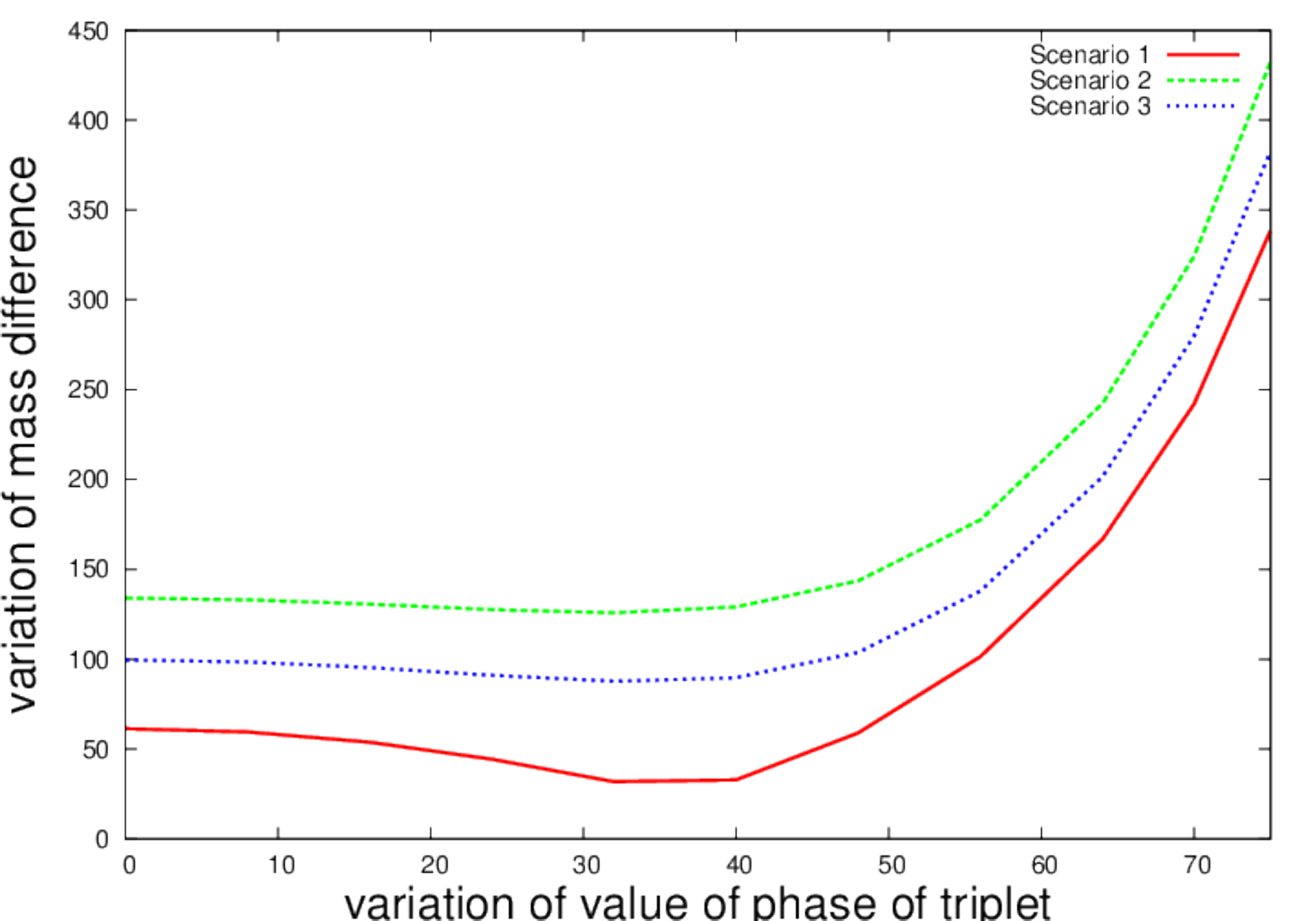}
\caption{}\label{fig:fig_g}
\end{subfigure}
\caption{Variation of mass difference between (\subref{fig:fig_f}) $\mathcal{H}_1^{++}$ and $\mathcal{H}_2^{++}$ and (\subref{fig:fig_g}) $\mathcal{H}_1^{++}$ and $\mathcal{H}_2^+$ with phase of triplet $\alpha$ 
}
\label{ac1}
\end{figure}

 One also notices a rather interesting effect of the triplet phase, shown in Figures~(\ref{fig:fig_f}) and~(\ref{fig:fig_g}). In Figure~(\ref{fig:fig_f}) we have plotted the variation of mass difference between $H_1^{++}$ and  $H_2^{++}$ with respect to variation of the phase $\alpha$. The mass difference between $H_1^{++}$ and  $H_2^+$ is similarly presented
 in Figure~(\ref{fig:fig_g}) . If  the mass differences do not allow the decays~(\ref{h50}) and~(\ref{hhw}) for $\alpha = 0$, they open up with  increase in
 the phase of the triplet, when all  other parameters are at fixed values.

 Earlier, we neglected contributions from the quartic terms in our scalar potential in the
 approximate forms of the doubly-and singly-charged mass matrices.  
 However, the import of the phase is not properly captured unless one
 retains these terms. Thus it is only via a full numerical analysis of
 the potential retaining all terms that the above effect of the phase of 
 the trilinear term becomes apparent.
 
 It should also be noted that the cosine of the complex phase suppresses
 the contribution to  neutrino masses. Consequently,  for the same 
 triplet vev, one requires larger values of the Yukawa interaction
 strengths. This makes the $l^+ l^+$ decay mode of a doubly charged
 scalar more competitive with $W^+ W^+ )$, as compared to the 
 results in~(\cite{avinanda}). 

 The branching ratios for a given scalar
in different channels are of course dependent on the various parameters that characterise a BP. We list all the charged scalar masses in Tables~\ref{charged1}, \ref{charged2} and~\ref{charged3}. Moreover, the neutral scalar masses are shown in Tables~\ref{neutral1}, \ref{neutral2} and~\ref{neutral3} for three different values of triplet phase $\alpha$. The  
branching ratios for $H^{++}_1$ and $H^{++}_2$ for different triplet phases are listed in Tables~\ref{branching1},\ref{branching2} and~\ref{branching3}, together with their pair-production cross sections at the LHC with $\sqrt{s} = 13$\,TeV. The cross sections and branching ratios have been calculated with the help of the  package FeynRules (version 1.6.0) \cite{neil,duhr}, 
thus creating a new UFO model file in MadGraph5-aMC@NLO (version 2.3.3)~\cite{mattelaer}. 
Using the full machinery of scalar mixing in this model, the decay widths into various channels 
have been obtained.

 The presence of the phase $\alpha$ also affects phenomenology in the following way. Suppose, in the absence of any additional symmetry in hierarchy, the matrices $B, E$ and $H$ are such that the resulting elements of each charged mass matrix are of similar order. This would normally result in rather low mass-splitting between $H_1^{++}$ and $H_2^{++}$, so long as the CP-violating phase is vanishing or small. For large $\alpha$, however, the degree of doublet-triplet mixing will be different for the two triplets. The splitting between $H_1^{++}$ and $H_2^{++}$ consequently goes up and the decay $H_1^{++} \rightarrow H_2^{++} h$ tends to dominate. Drell-Yan pair production of $H^{\pm \pm}_1$ is $h \ell^{\pm} \ell^{\pm}$, i.e an invariant mass peakin same-sign dileptons, together with an SM-like Higgs, that can be identified in the usual search channels at the LHC.

 From  Tables~\ref{branching1}, \ref{branching2} and~\ref{branching3}, we see that 
 decay~(\ref{hll2}) dominates, when  
 the masses of  $H_1^{++}$ and  $H_2^{++}$ are sufficiently separated. Also,
 when the phase space needed for this decay~(\ref{hll2}) is not available, the process~(\ref{hhw}) 
 dominates over all other remaining decays. Benchmark points when decay~(\ref{hhw}) mostly dominates for $H_1^{\pm \pm}$ have been discussed in detail in reference~\cite{avinanda}. Here we supplement those observations with some results for the case when decay~(\ref{hll2}) has an interesting consequence, as exemplitied by Figures~2 and~3.
 
\begin{figure}
\centering

\begin{subfigure}[t]{.4\textwidth}
\centering
\includegraphics[width=\linewidth]{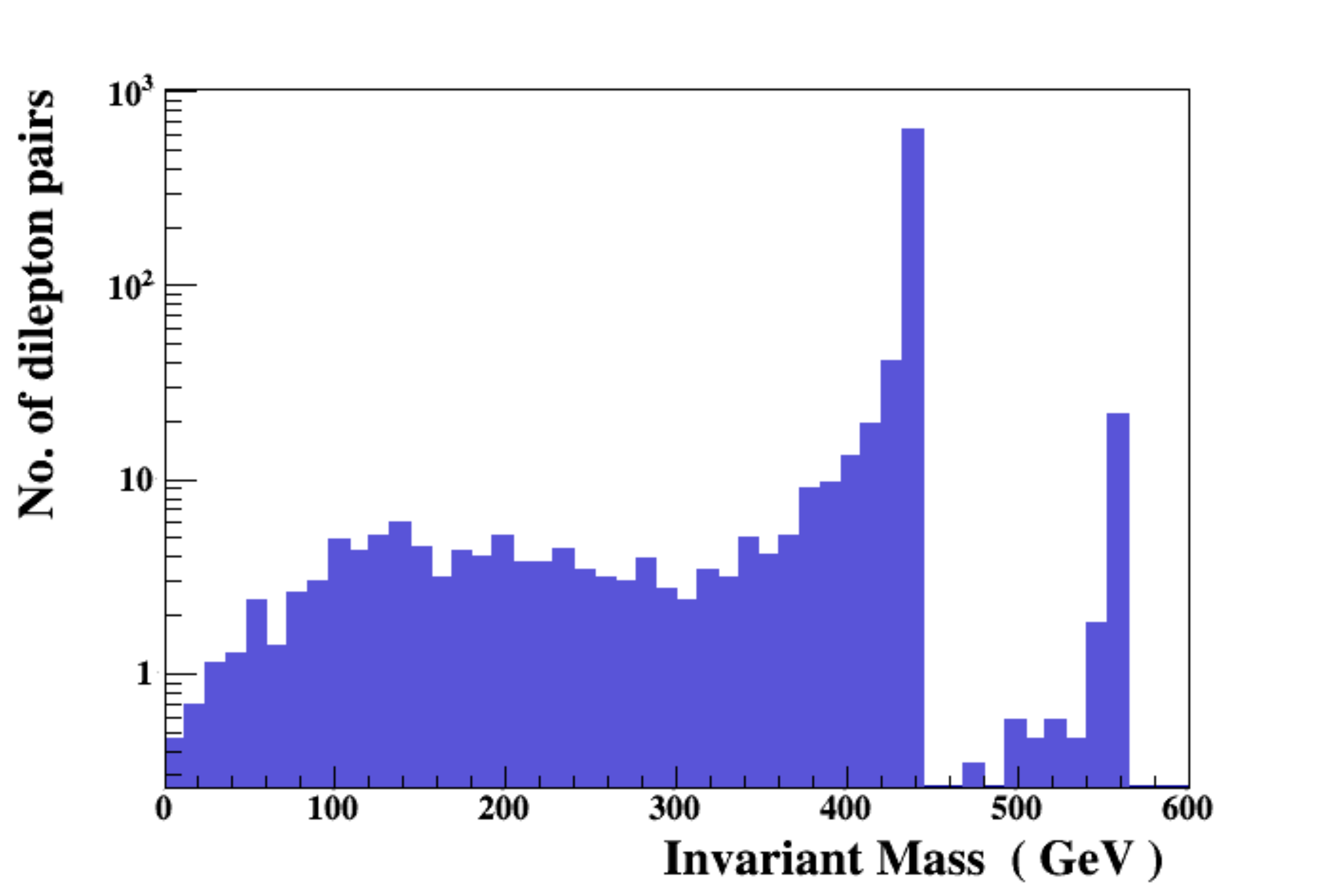}
        \caption{}\label{fig:fig_h}
\end{subfigure}
\begin{subfigure}[t]{.4\textwidth}
\centering
\includegraphics[width=\linewidth]{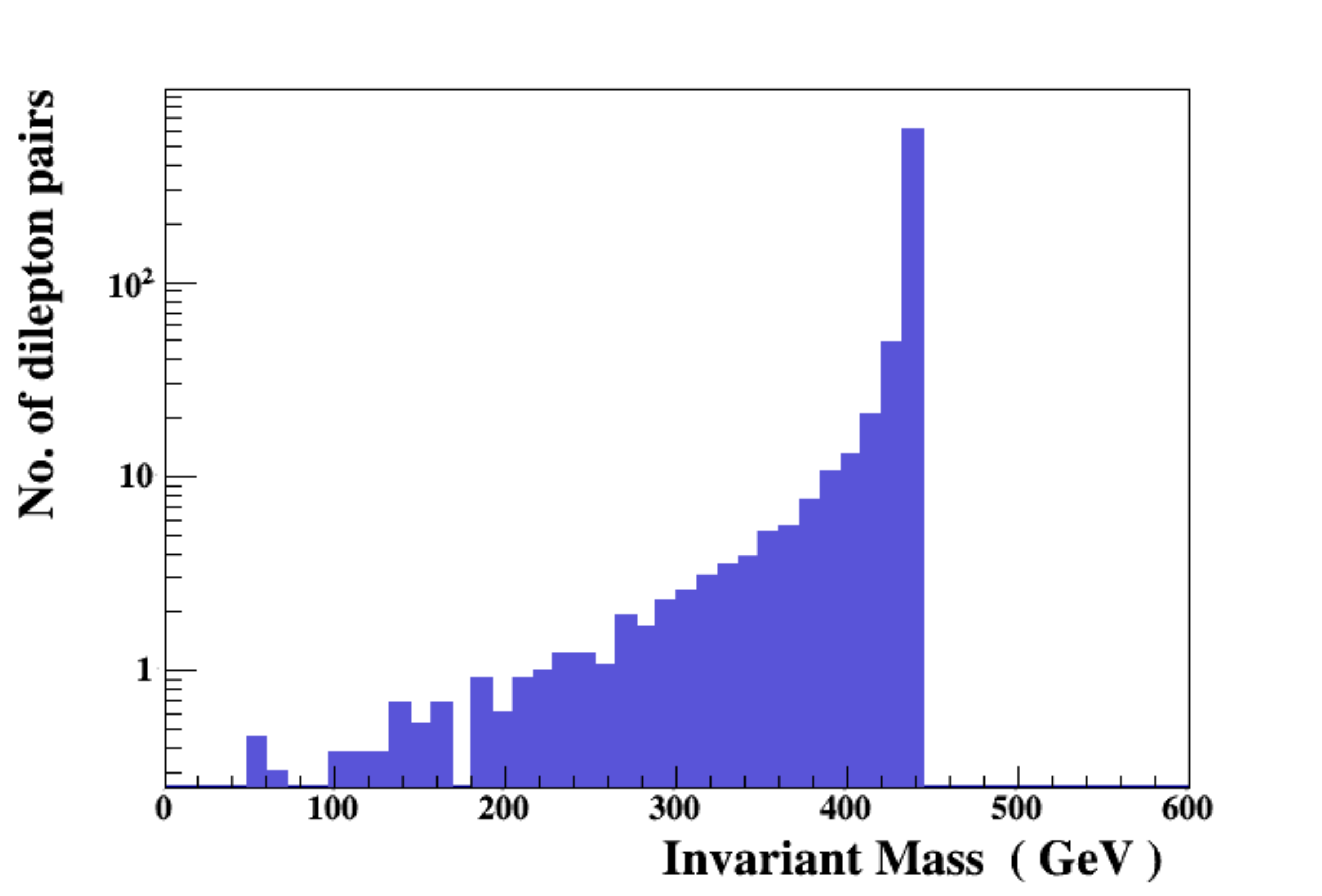}
\caption{}\label{fig:fig_i}
\end{subfigure}
\caption{Invariant mass distribution of same sign di-leptons for (\subref{fig:fig_h}) $\alpha = 60^{\circ}$ and (\subref{fig:fig_i}) $\alpha = 65^{\circ}$ for BP 4 of Scenario 2}
\label{ac2}
\end{figure}

 Figure~2 specifically shows the effect of CP-violating phase going up. We have seen in Figure~1 that the mass difference between $H_1^{++}$ and $H_2^{++}$ undergoes significant enhancement if, with other parameters unchanged, once the phase $\alpha$ is increased beyond $60^{\circ}$, the decay $H_1^{++} \rightarrow H_2^{++} h$ not only opens up but also becomes dominant. This point is accumulated in Figure~2, for which the choice of parameters is detailed in the caption. We simulated the same-sign dilepton final states and notice that there are two invariant mass peaks for $\alpha = 60^{\circ}$ in BP 4 of Scenario 2, as shown in Figure~(\ref{fig:fig_h}). However, Figure~(\ref{fig:fig_i}) shows only one peak for $\alpha = 65^{\circ}$, for which mass of $H_2^{++}$ remains practically the same. The increase in the CP-violating phase raises mass of $H_1^{++}$ and causes decay~(\ref{hll2}) to be overwhelmingly dominant. The two invariant mass peaks consequently make way for a single one at $M_{H_2^{++}}$. The identification of SM-like Higgs along with the same-sign dilepton peak can be an interesting signature of such a situation.
 
Figure~3 captures BP's, for which decay~(\ref{hhw}) is not overwhelmingly dominant. There, in addition to decay~(\ref{hhw}), process~(\ref{hll}), too has non-negligible branching ratios. For such situations, we have simulated the same-sign dilepton final states from both $H_1^{\pm \pm}$ and $H_2^{\pm \pm}$. \\  \noindent

%
 \noindent
 The leptons selected for this purpose satisfy: $\left| p_T^\mathrm{lepton} \right| > 20$ GeV, $|\eta_{\,\mathrm{lep}}| < 2.5$, $|\Delta R_{\,\mathrm{\ell \ell}}| < 0.2$ and $|\Delta R_{\,\mathrm{\ell j}}| < 0.4$ where $\Delta R^2 = \Delta \eta^2 + \Delta \phi^2$.
The two invariant mass peaks in each of the plots in Figure~3, bear testimony to the existence of two doubly-charged scalar states decaying in the dilepton channel.
\begin{figure}
\centering

\begin{subfigure}[t]{.4\textwidth}
\centering
\includegraphics[width=\linewidth]{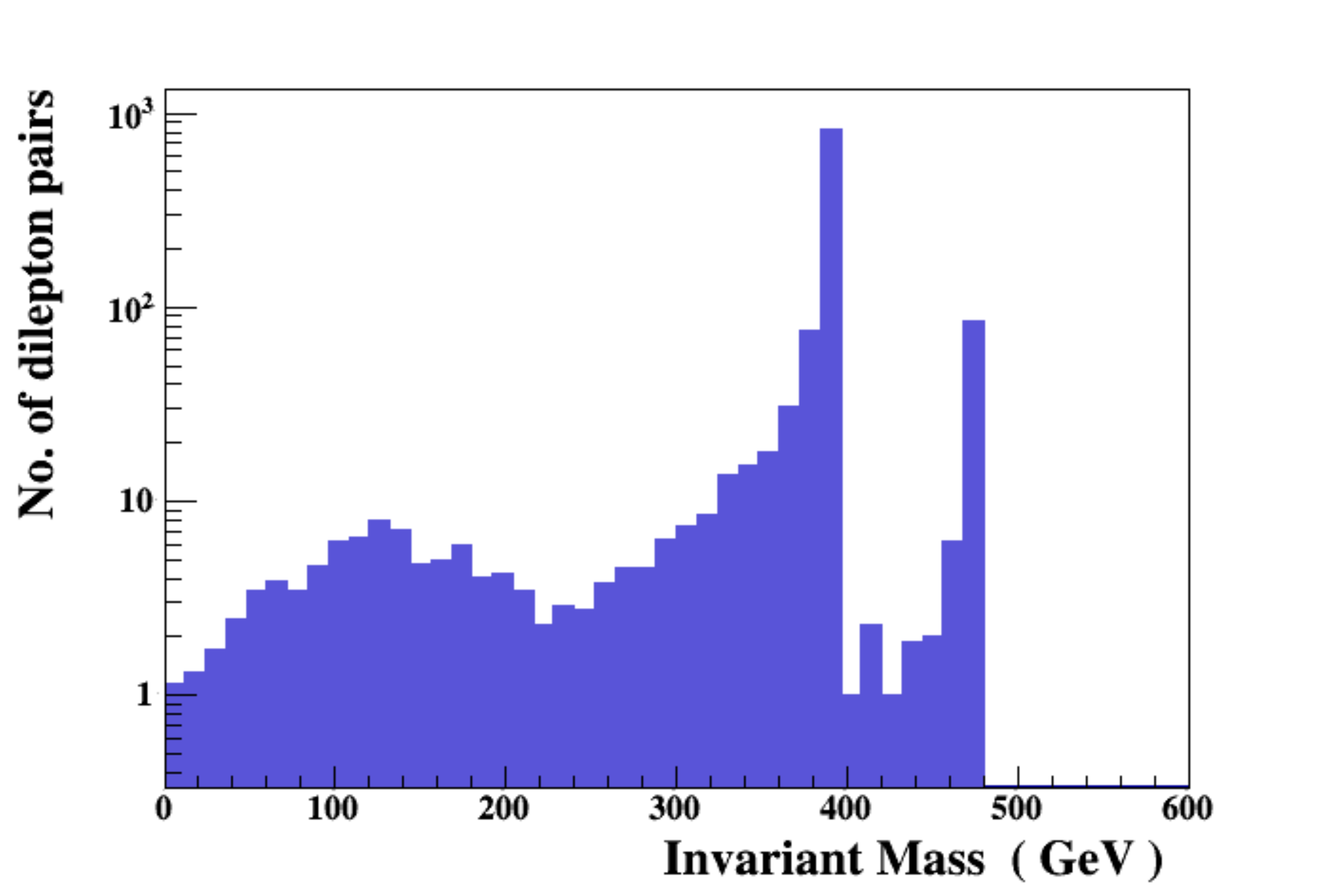}
        \caption{}\label{fig:fig_a}
\end{subfigure}
\begin{subfigure}[t]{.4\textwidth}
\centering
\includegraphics[width=\linewidth]{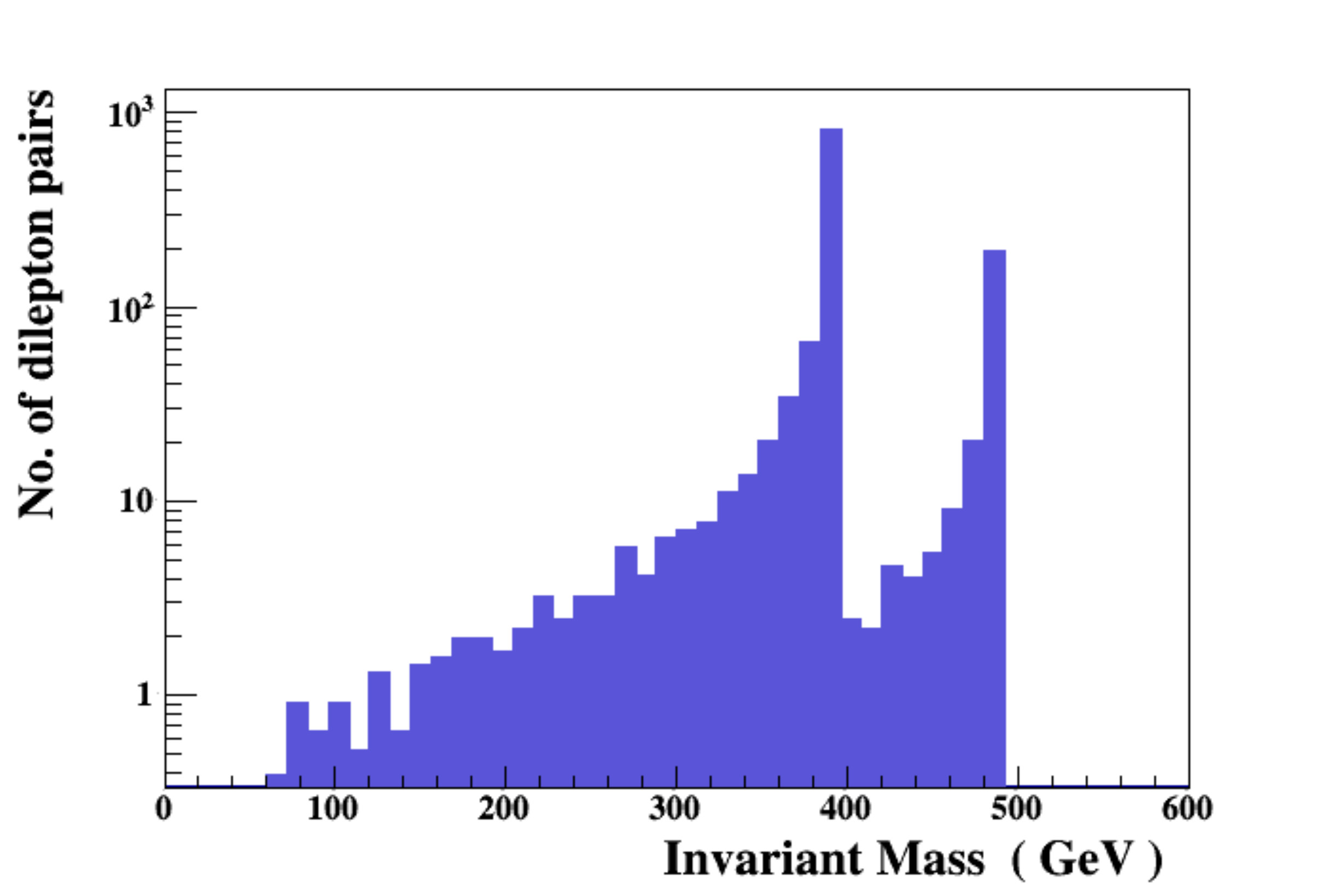}
\caption{}\label{fig:fig_b}
\end{subfigure}

\medskip

\begin{subfigure}[t]{.4\textwidth}
\centering
\includegraphics[width=\linewidth]{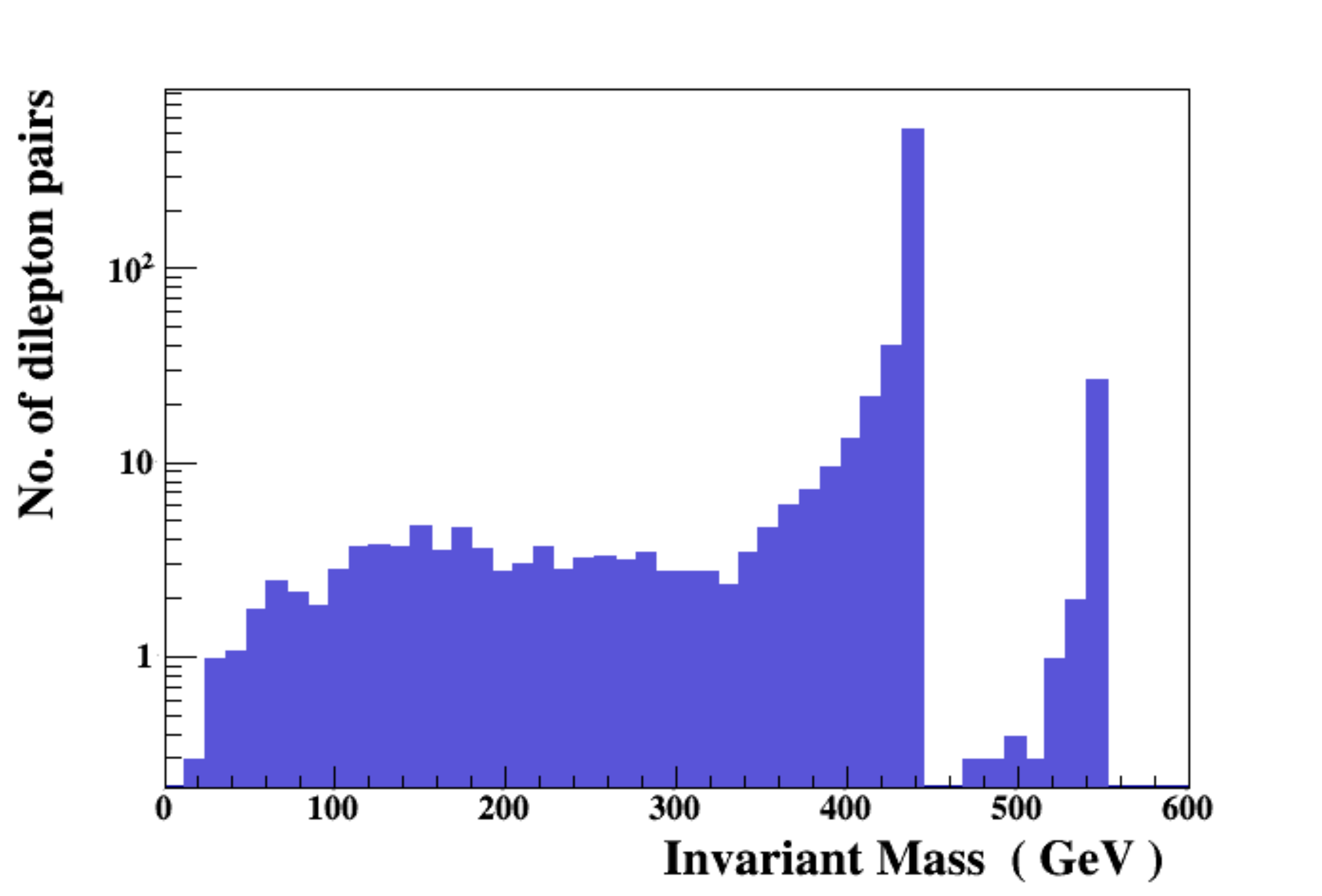}
        \caption{}\label{fig:fig_c}
\end{subfigure}
\begin{subfigure}[t]{.4\textwidth}
\centering
\includegraphics[width=\linewidth]{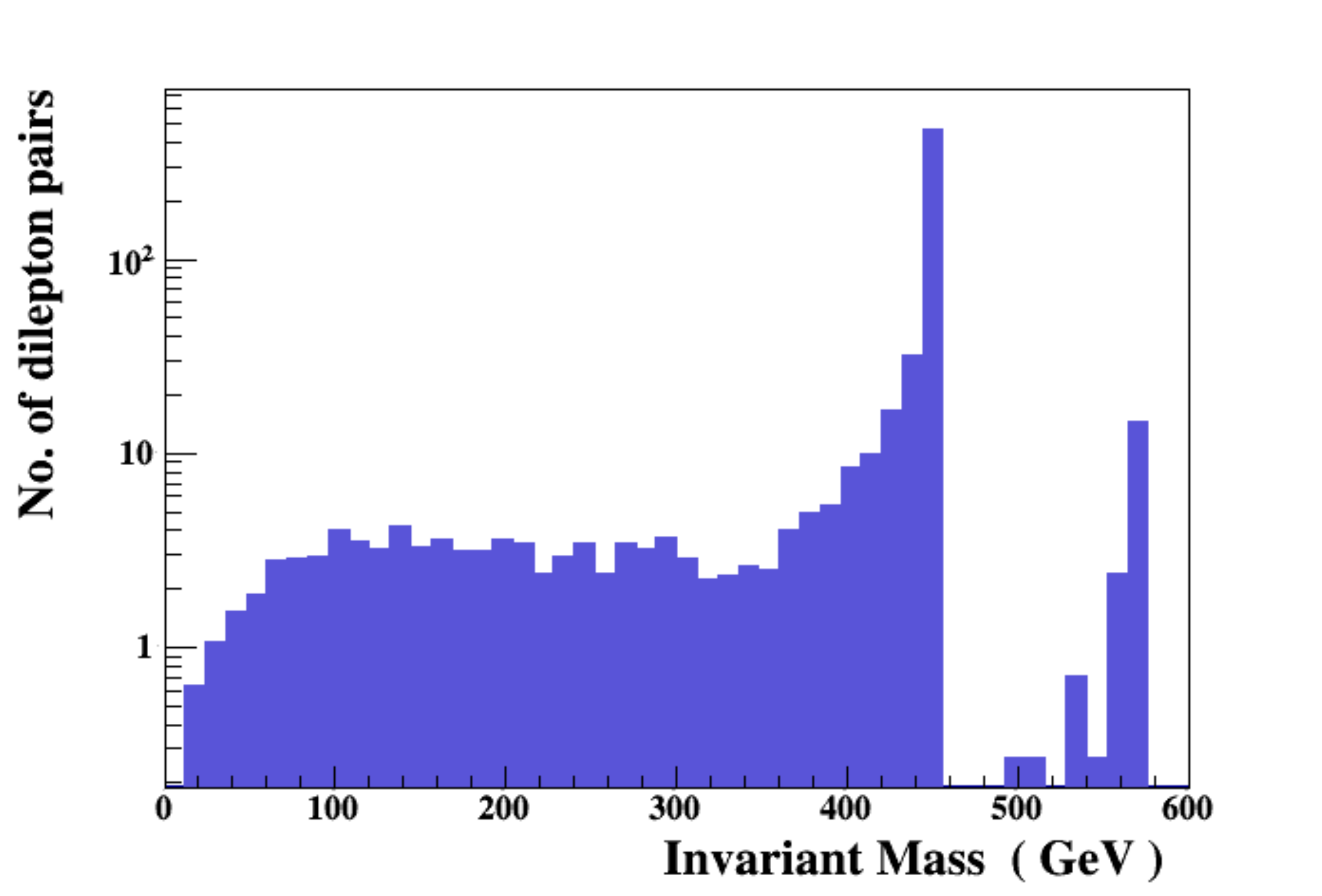}
\caption{}\label{fig:fig_d}
\end{subfigure}

\medskip

\begin{subfigure}[t]{.4\textwidth}
\centering
\vspace{0pt}
\includegraphics[width=\linewidth]{BP32.pdf}
\caption{}\label{fig:fig_e}
\end{subfigure}
\caption{Invariant mass distribution of same sign di-leptons for chosen benchmark points. In (\subref{fig:fig_a}) BP 3 and (\subref{fig:fig_b}) BP 4 of Scenario 2 for $\alpha = 30^{\circ}$. In (\subref{fig:fig_c}) BP 3 and (\subref{fig:fig_d}) BP 4 of Scenario 2 for $\alpha = 45^{\circ}$. In (\subref{fig:fig_e}) BP 4 of Scenario 2 for $\alpha = 60^{\circ}$}
\end{figure}
%
\begin{table}
\centering  
\begin{tabular}{|c||c|c|c|c|c|}

  \hline
   $\alpha = 30^{\circ}$ & Mass (GeV) & BP 1 & BP 2 & BP 3 & BP 4 \\ 
   \hline \hline
   & $H_1^{++}$ & $516.61$ & $513.83$ & $522.13$ & $537.62$
   \\ \cline{2-6} 
  Scenario 1 &  $H_2^{++}$ & $391.40$ & $389.82$ & $426.93$ &
   $440.96$ \\ \cline{2-6} 
   & $H_1^+$ & $516.58$ & $513.81$ & $522.10$ & $537.55$ \\ \cline{2-6} 
   & $H_2^+$ & $390.60$ & $389.79$ & $408.83$ & $416.30$ \\ 
  \hline
     & $H_1^{++}$ & $526.00$ & $529.85$ & $477.38$ & $485.76$
  \\ \cline{2-6} 
  Scenario 2 &  $H_2^{++}$ & $397.61$ & $390.10$ & $389.10$ &
  $389.33$ \\ \cline{2-6} 
   & $H_1^+$ & $525.94$ & $529.80$ & $477.34$ & $485.70$ \\ \cline{2-6} 
   & $H_2^+$ & $393.79$ & $390.01$ & $389.00$ & $389.28$ \\ 
  \hline
     & $H_1^{++}$ & $558.71$ & $562.35$ & $485.76$ & $477.38$
  \\ \cline{2-6} 
  Scenario 3 &  $H_2^{++}$ & $427.00$ & $407.20$ & $389.33$ &
  $389.10$ \\ \cline{2-6} 
   & $H_1^+$ & $557.59$ & $559.11$ & $485.70$ & $477.34$ \\ \cline{2-6} 
   & $H_2^+$ & $392.90$ & $405.91$ & $389.28$ & $389.00$ \\ 
  \hline
\end{tabular}
\caption{Charged scalar masses for phase $\alpha = 30^{\circ}$. \label{charged1}}  
\end{table}
\clearpage
\begin{table}
\centering  
\begin{tabular}{|c||c|c|c|c|c|}

  \hline
   $\alpha = 30^{\circ}$ & Mass (GeV) & BP 1 & BP 2 & BP 3 & BP 4 \\ 
   \hline \hline
   & $H_{01}$ & $730.57$ & $726.65$ & $738.36$ & $760.19$
   \\ \cline{2-6} 
  Scenario 1 &  $H_{02}$ & $730.49$ & $726.62$ & $738.32$ &
   $760.15$ \\ \cline{2-6} 
   & $H_{03}$ & $552.30$ & $551.25$ & $551.40$ & $552.65$ \\ \cline{2-6} 
   & $H_{04}$ & $552.15$ & $551.20$ & $551.34$ & $552.56$ \\ \cline{2-6} 
   & $h$ & $125.16$ & $125.18$ & $125.20$ & $125.15$ \\
  \hline
     & $H_{01}$ & $743.85$ & $749.33$ & $675.11$ & $686.96$
  \\ \cline{2-6} 
  Scenario 2 &  $H_{02}$ & $743.00$ & $749.25$ & $675.00$ &
  $686.90$ \\ \cline{2-6} 
   & $H_{03}$ & $552.21$ & $551.50$ & $550.17$ & $550.53$ \\ \cline{2-6} 
   & $H_{04}$ & $552.10$ & $551.39$ & $550.05$ & $550.40$ \\ \cline{2-6} 
   & $h$ & $125.21$ & $125.18$ & $125.22$ & $125.23$ \\
  \hline
     & $H_{01}$ & $787.10$ & $789.25$ & $687.00$ & $676.15$
  \\ \cline{2-6} 
  Scenario 3 &  $H_{02}$ & $787.00$ & $789.10$ & $686.90$ &
  $676.08$ \\ \cline{2-6} 
   & $H_{03}$ & $541.16$ & $542.00$ & $552.21$ & $549.90$ \\ \cline{2-6} 
   & $H_{04}$ & $541.07$ & $541.91$ & $552.00$ & $549.75$ \\ \cline{2-6} 
   & $h$ & $125.23$ & $125.26$ & $125.13$ & $125.16$ \\
  \hline
\end{tabular}
\caption{Neutral scalar masses for phase $\alpha = 30^{\circ}$. \label{neutral1}}  
\end{table}
\clearpage
\begin{table}
\centering  
\begin{tabular}{|c||c|c|c|c|c|}
  \hline
   $\alpha = 30^{\circ}$ & Data & BP 1 & BP 2 & BP 3 & BP 4 \\ 
   \hline \hline
 & $\mbox{BR}(H_1^{++} \rightarrow H_2^{++} h)$&$5.1 \times10^{-3}$&not allowed&not allowed&not allowed \\ \cline{2-6} 
 & $\mbox{BR}(H_1^{++} \rightarrow H_2^+ W^+)$&$0.99 $
   &$0.99$&$0.79$&$0.99$\\ \cline{2-6} 
& $\mbox{BR}(H_1^{++} \rightarrow W^+
   W^+)$&$2.8 \times 10^{-3}$&$6.5 \times 10^{-2}$&$0.21$&$3.1 \times 10^{-7}$\\ \cline{2-6} 
& $\mbox{BR}(H_1^{++} \rightarrow \ell^+_i \ell^+_j)$&$4.8 \times 10^{-21}$&$1.6
   \times 10^{-20}$&$1.3 \times 10^{-18}$&$2.1 \times
   10^{-23}$\\ \cline{2-6} 
Scenario 1 & $\mbox{BR}(H_2^{++} \rightarrow W^+
   W^+)$&$0.99$&$0.99$&$0.99$&$0.99$\\ \cline{2-6} 
& $\mbox{BR}(H_2^{++} \rightarrow \ell^+_i \ell^+_j)$&$1.6 \times 10^{-18}$&$2.7
   \times 10^{-18}$&$3.9 \times 10^{-17}$&$1.3 \times
   10^{-20}$\\ \cline{2-6} 
 & $\sigma(pp \rightarrow H_1^{++} H_1^{--})$&$\ 1.10$\,fb
&$\ 1.13$\,fb&$\ 1.05$\,fb&$\ 0.42$\,fb\\ \cline{2-6}
& $\sigma(pp \rightarrow H_2^{++} H_2^{--})$&$\ 3.97$\,fb&$\ 4.10
  $\,fb&$\ 2.70$\,fb&$\ 1.06$\,fb\\ 
   \hline
 & $\mbox{BR}(H_1^{++} \rightarrow H_2^{++} h)$&$0.84$&$0.96$&not allowed&not allowed \\ \cline{2-6} 
 & $\mbox{BR}(H_1^{++} \rightarrow H_2^+
   W^+)$&$0.13$&$0.03$&$0.76$&$0.42$\\ \cline{2-6} 
 & $\mbox{BR}(H_1^{++} \rightarrow W^+ W^+)$&$3.1 \times 10^{-20}$&$1.9
   \times 10^{-20}$&$1.2 \times 10^{-19}$&$2.8 \times
   10^{-21}$\\ \cline{2-6} 
 & $\mbox{BR}(H_1^{++} \rightarrow \ell^+_i \ell^+_j)$ 
&$0.03$&$8.9 \times10^{-3}$&$0.24$&$0.58$\\ \cline{2-6} 
Scenario 2 & $\mbox{BR}(H_2^{++} \rightarrow W^+ W^+)$&$1.9 \times
10^{-19}$&$2.8 \times 10^{-19}$&$1.8 \times 10^{-20}$&$3.7 \times
10^{-19}$\\ \cline{2-6} 
& $\mbox{BR}(H_2^{++} \rightarrow \ell^+_i \ell^+_j)$
&$0.99$&$0.99$&$0.99$&$0.99$\\ \cline{2-6} 
& $\sigma(pp \rightarrow H_1^{++} H_1^{--})$&$\ 1.01$\,fb&
$\ 1.02$\,fb&$\ 1.56$\,fb&$\ 1.43$\,fb\\ \cline{2-6} 
& $\sigma(pp \rightarrow H_2^{++} H_2^{--})$&$\ 3.60$\,fb&
$\ 4.04$\,fb&$\ 3.97$\,fb&$\ 3.95$\,fb\\ 
  \hline
& $\mbox{BR}(H_1^{++} \rightarrow H_2^{++} h)$&$0.99$&$0.99$&not allowed&not allowed \\ \cline{2-6} 
& $\mbox{BR}(H_1^{++} \rightarrow H_2^+
  W^+)$&$2.1 \times 10^{-3}$&$1.3 \times 10^{-2}$&$0.99$&$0.99$\\ \cline{2-6} 
& $\mbox{BR}(H_1^{++} \rightarrow W^+ W^+)$&$2.6 \times 10^{-14}$&$3.1
  \times 10^{-14}$&$4.3 \times 10^{-10}$&$2.8 \times
  10^{-11}$\\ \cline{2-6} 
& $\mbox{BR}(H_1^{++} \rightarrow \ell^+_i \ell^+_j)$&$1.5 \times 10^{-11}$&$2.3
  \times 10^{-11}$&$3.7 \times 10^{-7}$&$5.4 \times
  10^{-8}$\\ \cline{2-6} 
Scenario 3 & $\mbox{BR}(H_2^{++} \rightarrow W^+ W^+)$&$0.03
$&$0.01$&$0.04$&$0.02$\\ \cline{2-6} 
& $\mbox{BR}(H_2^{++} \rightarrow \ell^+_i \ell^+_j)$
&$0.97$&$0.99$&$0.96$&$0.98$\\ \cline{2-6} 
& $\sigma(pp \rightarrow H_1^{++} H_1^{--})$&$\ 0.77$\,fb&
$\ 0.74$\,fb&$\ 1.45$\,fb&$\ 1.58$\,fb\\ \cline{2-6} 
& $\sigma(pp \rightarrow H_2^{++} H_2^{--})$&$\ 3.61$\,fb&
$\ 2.75$\,fb&$\ 3.95$\,fb&$\ 3.98$\,fb\\ 
  \hline
\end{tabular}
\caption{Decay branching ratios and production cross sections for
  doubly-charged scalars for phase $\alpha = 30^{\circ}$. \label{branching1}} 
\end{table}
\clearpage
\begin{table}
\centering  
\begin{tabular}{|c||c|c|c|c|c|}

  \hline
   $\alpha = 45^{\circ}$ & Mass (GeV) & BP 1 & BP 2 & BP 3 & BP 4 \\ 
   \hline \hline
   & $H_1^{++}$ & $542.27$ & $539.35$ & $549.85$ & $566.55$
   \\ \cline{2-6} 
  Scenario 1 &  $H_2^{++}$ & $406.60$ & $405.20$ & $438.46$ &
   $450.96$ \\ \cline{2-6} 
   & $H_1^+$ & $542.22$ & $539.20$ & $548.73$ & $564.92$ \\ \cline{2-6} 
   & $H_2^+$ & $405.90$ & $405.07$ & $422.28$ & $428.94$ \\ 
  \hline
     & $H_1^{++}$ & $543.30$ & $538.15$ & $551.62$ & $564.34$
  \\ \cline{2-6} 
  Scenario 2 &  $H_2^{++}$ & $405.10$ & $404.10$ & $440.10$ &
  $448.82$ \\ \cline{2-6} 
   & $H_1^+$ & $542.50$ & $537.65$ & $550.90$ & $563.65$ \\ \cline{2-6} 
   & $H_2^+$ & $405.00$ & $403.20$ & $439.72$ & $447.90$ \\ 
  \hline
     & $H_1^{++}$ & $545.82$ & $540.32$ & $550.90$ & $567.80$
  \\ \cline{2-6} 
  Scenario 3 &  $H_2^{++}$ & $409.80$ & $405.00$ & $439.50$ &
  $452.45$ \\ \cline{2-6} 
   & $H_1^+$ & $544.71$ & $539.46$ & $550.15$ & $565.90$ \\ \cline{2-6} 
   & $H_2^+$ & $409.00$ & $404.75$ & $425.38$ & $432.80$ \\ 
  \hline
\end{tabular}
\caption{Charged scalar masses for phase $\alpha = 45^{\circ}$. \label{charged2}}  
\end{table}
\clearpage
\begin{table}
\centering  
\begin{tabular}{|c||c|c|c|c|c|}

  \hline
   $\alpha = 45^{\circ}$ & Mass (GeV) & BP 1 & BP 2 & BP 3 & BP 4 \\ 
   \hline \hline
   & $H_{01}$ & $766.78$ & $762.75$ & $774.79$ & $797.22$
   \\ \cline{2-6} 
  Scenario 1 &  $H_{02}$ & $766.60$ & $762.11$ & $774.23$ &
   $797.00$ \\ \cline{2-6} 
   & $H_{03}$ & $573.00$ & $572.50$ & $575.50$ & $576.17$ \\ \cline{2-6} 
   & $H_{04}$ & $572.65$ & $572.00$ & $575.15$ & $575.95$ \\ \cline{2-6} 
   & $h$ & $125.15$ & $125.22$ & $125.19$ & $125.13$ \\
  \hline
     & $H_{01}$ & $768.10$ & $760.37$ & $772.90$ & $795.85$
  \\ \cline{2-6} 
  Scenario 2 &  $H_{02}$ & $768.00$ & $760.13$ & $772.75$ &
  $795.50$ \\ \cline{2-6} 
   & $H_{03}$ & $575.32$ & $570.00$ & $574.30$ & $576.85$ \\ \cline{2-6} 
   & $H_{04}$ & $575.15$ & $569.22$ & $573.78$ & $576.20$ \\ \cline{2-6} 
   & $h$ & $125.12$ & $125.16$ & $125.24$ & $125.17$ \\
  \hline
     & $H_{01}$ & $771.10$ & $758.52$ & $778.10$ & $798.37$
  \\ \cline{2-6} 
  Scenario 3 &  $H_{02}$ & $770.85$ & $758.00$ & $777.85$ &
  $798.00$ \\ \cline{2-6} 
   & $H_{03}$ & $577.31$ & $568.75$ & $578.29$ & $577.21$ \\ \cline{2-6} 
   & $H_{04}$ & $577.00$ & $568.13$ & $578.00$ & $577.00$ \\ \cline{2-6} 
   & $h$ & $125.18$ & $125.21$ & $125.13$ & $125.16$ \\
  \hline
\end{tabular}
\caption{Neutral scalar masses for phase $\alpha = 45^{\circ}$. \label{neutral2}}  
\end{table}
\clearpage
\begin{table}
\centering  
\begin{tabular}{|c||c|c|c|c|c|}
  \hline
   $\alpha = 45^{\circ}$ & Data & BP 1 & BP 2 & BP 3 & BP 4 \\ 
   \hline \hline
 & $\mbox{BR}(H_1^{++} \rightarrow H_2^{++} h)$&$0.99$&$0.99$&not allowed&not allowed \\ \cline{2-6} 
 & $\mbox{BR}(H_1^{++} \rightarrow H_2^+ W^+)$&$8.2 \times 10^{-4} $
   &$9.1 \times 10^{-4}$&$0.90$&$0.96$\\ \cline{2-6} 
& $\mbox{BR}(H_1^{++} \rightarrow W^+
   W^+)$&$2.4 \times 10^{-5}$&$1.7 \times 10^{-4}$&$0.09$&$0.04$\\ \cline{2-6} 
& $\mbox{BR}(H_1^{++} \rightarrow \ell^+_i \ell^+_j)$&$3.1 \times 10^{-22}$&$3.8
   \times 10^{-22}$&$6.1 \times 10^{-19}$&$4.2 \times
   10^{-20}$\\ \cline{2-6} 
Scenario 1 & $\mbox{BR}(H_2^{++} \rightarrow W^+
   W^+)$&$0.99$&$0.99$&$0.99$&$0.99$\\ \cline{2-6} 
& $\mbox{BR}(H_2^{++} \rightarrow \ell^+_i \ell^+_j)$&$7.4 \times 10^{-19}$&$8.3begin
   \times 10^{-19}$&$2.1 \times 10^{-19}$&$6.7 \times
   10^{-19}$\\ \cline{2-6} 
 & $\sigma(pp \rightarrow H_1^{++} H_1^{--})$&$\ 0.88$\,fb
&$\ 0.87$\,fb&$\ 0.80$\,fb&$\ 0.71$\,fb\\ \cline{2-6}
& $\sigma(pp \rightarrow H_2^{++} H_2^{--})$&$\ 3.39$\,fb&$\ 3.33
  $\,fb&$\ 2.43$\,fb&$\ 2.13$\,fb\\ 
   \hline
 & $\mbox{BR}(H_1^{++} \rightarrow H_2^{++} h)$&$0.99$&$0.99$&not allowed&not allowed \\ \cline{2-6} 
 & $\mbox{BR}(H_1^{++} \rightarrow H_2^+
   W^+)$&$4.5 \times 10^{-4}$&$3.9 \times 10^{-4}$&$0.64$&$0.88$\\ \cline{2-6} 
 & $\mbox{BR}(H_1^{++} \rightarrow W^+ W^+)$&$1.3 \times 10^{-21}$&$9.1
   \times 10^{-22}$&$7.6 \times 10^{-20}$&$6.4 \times
   10^{-21}$\\ \cline{2-6} 
 & $\mbox{BR}(H_1^{++} \rightarrow \ell^+_i \ell^+_j)$ 
&$3.1 \times 10^{-6}$&$1.7 \times10^{-4}$&$0.36$&$0.12$\\ \cline{2-6} 
Scenario 2 & $\mbox{BR}(H_2^{++} \rightarrow W^+ W^+)$&$2.7 \times
10^{-20}$&$2.5 \times 10^{-19}$&$3.2 \times 10^{-19}$&$5.7 \times
10^{-20}$\\ \cline{2-6} 
& $\mbox{BR}(H_2^{++} \rightarrow \ell^+_i \ell^+_j)$
&$0.99$&$0.99$&$0.99$&$0.99$\\ \cline{2-6} 
& $\sigma(pp \rightarrow H_1^{++} H_1^{--})$&$\ 0.93$\,fb&
$\ 0.89$\,fb&$\ 0.86$\,fb&$\ 0.75$\,fb\\ \cline{2-6} 
& $\sigma(pp \rightarrow H_2^{++} H_2^{--})$&$\ 2.55$\,fb&
$\ 3.35$\,fb&$\ 2.46$\,fb&$\ 2.18$\,fb\\ 
  \hline
& $\mbox{BR}(H_1^{++} \rightarrow H_2^{++} h)$&$0.99$&$0.99$&not allowed&not allowed \\ \cline{2-6} 
& $\mbox{BR}(H_1^{++} \rightarrow H_2^+
  W^+)$&$3.6 \times 10^{-5}$&$1.4 \times 10^{-4}$&$0.99$&$0.99$\\ \cline{2-6} 
& $\mbox{BR}(H_1^{++} \rightarrow W^+ W^+)$&$8.6 \times 10^{-14}$&$7.3
  \times 10^{-13}$&$1.4 \times 10^{-9}$&$5.8 \times
  10^{-10}$\\ \cline{2-6} 
& $\mbox{BR}(H_1^{++} \rightarrow \ell^+_i \ell^+_j)$&$4.8 \times 10^{-11}$&$3.7
  \times 10^{-11}$&$5.6 \times 10^{-11}$&$4.7 \times
  10^{-9}$\\ \cline{2-6} 
Scenario 3 & $\mbox{BR}(H_2^{++} \rightarrow W^+ W^+)$&$0.02
$&$0.04$&$0.97$&$0.05$\\ \cline{2-6} 
& $\mbox{BR}(H_2^{++} \rightarrow \ell^+_i \ell^+_j)$
&$0.98$&$0.96$&$0.03$&$0.95$\\ \cline{2-6} 
& $\sigma(pp \rightarrow H_1^{++} H_1^{--})$&$\ 0.92$\,fb&
$\ 0.95$\,fb&$\ 0.84$\,fb&$\ 0.73$\,fb\\ \cline{2-6} 
& $\sigma(pp \rightarrow H_2^{++} H_2^{--})$&$\ 3.42$\,fb&
$\ 3.37$\,fb&$\ 2.44$\,fb&$\ 2.15$\,fb\\ 
  \hline
\end{tabular}
\caption{Decay branching ratios and production cross sections for
  doubly-charged scalars for phase $\alpha = 45^{\circ}$. \label{branching2}} 
\end{table}
\clearpage
\begin{table}
\centering  
\begin{tabular}{|c||c|c|c|c|c|}

  \hline
   $\alpha = 60^{\circ}$ & Mass (GeV) & BP 1 & BP 2 & BP 3 & BP 4 \\ 
   \hline \hline
   & $H_1^{++}$ & $557.90$ & $563.51$ & $564.20$ & $556.56$
   \\ \cline{2-6} 
  Scenario 1 &  $H_2^{++}$ & $412.20$ & $411.51$ & $434.37$ &
   $439.71$ \\ \cline{2-6} 
   & $H_1^+$ & $557.62$ & $563.25$ & $559.18$ & $548.00$ \\ \cline{2-6} 
   & $H_2^+$ & $411.65$ & $411.18$ & $423.27$ & $426.15$ \\ 
  \hline
     & $H_1^{++}$ & $558.20$ & $565.20$ & $566.40$ & $554.30$
  \\ \cline{2-6} 
  Scenario 2 &  $H_2^{++}$ & $411.90$ & $413.61$ & $436.56$ &
  $438.12$ \\ \cline{2-6} 
   & $H_1^+$ & $558.00$ & $564.50$ & $565.90$ & $553.65$ \\ \cline{2-6} 
   & $H_2^+$ & $410.75$ & $412.85$ & $435.85$ & $435.32$ \\ 
  \hline
     & $H_1^{++}$ & $556.65$ & $560.30$ & $567.80$ & $552.90$
  \\ \cline{2-6} 
  Scenario 3 &  $H_2^{++}$ & $410.25$ & $408.35$ & $437.90$ &
  $436.59$ \\ \cline{2-6} 
   & $H_1^+$ & $556.00$ & $559.75$ & $563.21$ & $550.00$ \\ \cline{2-6} 
   & $H_2^+$ & $409.85$ & $407.80$ & $425.56$ & $429.11$ \\ 
  \hline
\end{tabular}
\caption{Charged scalar masses for phase $\alpha = 60^{\circ}$. \label{charged3}}  
\end{table}
\clearpage
\begin{table}
\centering  
\begin{tabular}{|c||c|c|c|c|c|}

  \hline
   $\alpha = 60^{\circ}$ & Mass (GeV) & BP 1 & BP 2 & BP 3 & BP 4 \\ 
   \hline \hline
   & $H_{01}$ & $788.52$ & $796.91$ & $784.64$ & $765.05$
   \\ \cline{2-6} 
  Scenario 1 &  $H_{02}$ & $788.35$ & $796.27$ & $784.21$ &
   $764.62$ \\ \cline{2-6} 
   & $H_{03}$ & $581.43$ & $583.16$ & $579.62$ & $577.78$ \\ \cline{2-6} 
   & $H_{04}$ & $581.32$ & $582.95$ & $579.13$ & $577.21$ \\ \cline{2-6} 
   & $h$ & $125.16$ & $125.24$ & $125.14$ & $125.20$ \\
  \hline
     & $H_{01}$ & $790.21$ & $793.82$ & $786.52$ & $762.90$
  \\ \cline{2-6} 
  Scenario 2 &  $H_{02}$ & $790.00$ & $793.11$ & $786.09$ &
  $762.42$ \\ \cline{2-6} 
   & $H_{03}$ & $579.32$ & $580.16$ & $582.32$ & $574.21$ \\ \cline{2-6} 
   & $H_{04}$ & $579.00$ & $579.92$ & $582.00$ & $573.86$ \\ \cline{2-6} 
   & $h$ & $125.15$ & $125.10$ & $125.21$ & $125.09$ \\
  \hline
     & $H_{01}$ & $786.51$ & $790.63$ & $785.00$ & $760.71$
  \\ \cline{2-6} 
  Scenario 3 &  $H_{02}$ & $786.00$ & $790.27$ & $784.32$ &
  $760.29$ \\ \cline{2-6} 
   & $H_{03}$ & $577.82$ & $576.21$ & $580.14$ & $570.90$ \\ \cline{2-6} 
   & $H_{04}$ & $577.50$ & $576.00$ & $579.55$ & $570.58$ \\ \cline{2-6} 
   & $h$ & $125.23$ & $125.14$ & $125.23$ & $125.18$ \\
  \hline
\end{tabular}
\caption{Neutral scalar masses for phase $\alpha = 60^{\circ}$. \label{neutral3}}  
\end{table}
\clearpage
\begin{table}
\centering  
\begin{tabular}{|c||c|c|c|c|c|}
  \hline
   $\alpha = 60^{\circ}$ & Data & BP 1 & BP 2 & BP 3 & BP 4 \\ 
   \hline \hline
 & $\mbox{BR}(H_1^{++} \rightarrow H_2^{++} h)$&$0.99$&$0.98$&$0.99$&not allowed \\ \cline{2-6} 
 & $\mbox{BR}(H_1^{++} \rightarrow H_2^+ W^+)$&$3.9 \times 10^{-4} $
   &$2.6 \times 10^{-2}$&$0.01$&$0.94$\\ \cline{2-6} 
& $\mbox{BR}(H_1^{++} \rightarrow W^+
   W^+)$&$1.7 \times 10^{-5}$&$8.9 \times 10^{-3}$&$6.9 \times 10^{-5}$&$0.05$\\ \cline{2-6} 
& $\mbox{BR}(H_1^{++} \rightarrow \ell^+_i \ell^+_j)$&$2.6 \times 10^{-22}$&$4.7
   \times 10^{-21}$&$3.1 \times 10^{-22}$&$5.6 \times
   10^{-20}$\\ \cline{2-6} 
Scenario 1 & $\mbox{BR}(H_2^{++} \rightarrow W^+
   W^+)$&$0.99$&$0.99$&$0.99$&$0.99$\\ \cline{2-6} 
& $\mbox{BR}(H_2^{++} \rightarrow \ell^+_i \ell^+_j)$&$6.7 \times 10^{-19}$&$7.2
   \times 10^{-19}$&$5.3 \times 10^{-20}$&$3.5 \times
   10^{-15}$\\ \cline{2-6} 
 & $\sigma(pp \rightarrow H_1^{++} H_1^{--})$&$\ 0.79$\,fb
&$\ 0.71$\,fb&$\ 0.72$\,fb&$\ 0.84$\,fb\\ \cline{2-6}
& $\sigma(pp \rightarrow H_2^{++} H_2^{--})$&$\ 3.22$\,fb&$\ 3.15
  $\,fb&$\ 2.48$\,fb&$\ 2.51$\,fb\\ 
   \hline
 & $\mbox{BR}(H_1^{++} \rightarrow H_2^{++} h)$&$0.99$&$0.79$&$0.99$&not allowed \\ \cline{2-6} 
 & $\mbox{BR}(H_1^{++} \rightarrow H_2^+
   W^+)$&$3.2 \times 10^{-5}$&$0.21$&$3.1 \times 10^{-3}$&$0.88$\\ \cline{2-6} 
 & $\mbox{BR}(H_1^{++} \rightarrow W^+ W^+)$&$3.9 \times 10^{-22}$&$5.6
   \times 10^{-22}$&$5.9 \times 10^{-22}$&$4.3 \times
   10^{-22}$\\ \cline{2-6} 
 & $\mbox{BR}(H_1^{++} \rightarrow \ell^+_i \ell^+_j)$ 
&$2.1 \times 10^{-4}$&$1.4 \times10^{-4}$&$3.7 \times10^{-4}$&$0.12$\\ \cline{2-6} 
Scenario 2 & $\mbox{BR}(H_2^{++} \rightarrow W^+ W^+)$&$3.2 \times
10^{-19}$&$5.6 \times 10^{-20}$&$7.8 \times 10^{-18}$&$6.3 \times
10^{-19}$\\ \cline{2-6} 
& $\mbox{BR}(H_2^{++} \rightarrow \ell^+_i \ell^+_j)$
&$0.99$&$0.99$&$0.99$&$0.99$\\ \cline{2-6} 
& $\sigma(pp \rightarrow H_1^{++} H_1^{--})$&$\ 0.77$\,fb&
$\ 0.74$\,fb&$\ 0.81$\,fb&$\ 0.86$\,fb\\ \cline{2-6} 
& $\sigma(pp \rightarrow H_2^{++} H_2^{--})$&$\ 3.26$\,fb&
$\ 3.19$\,fb&$\ 2.46$\,fb&$\ 2.53$\,fb\\ 
  \hline
& $\mbox{BR}(H_1^{++} \rightarrow H_2^{++} h)$&$0.99$&$0.90$&$0.99$&not allowed \\ \cline{2-6} 
& $\mbox{BR}(H_1^{++} \rightarrow H_2^+
  W^+)$&$2.5 \times 10^{-4}$&$0.10$&$1.2 \times 10^{-2}$&$0.99$\\ \cline{2-6} 
& $\mbox{BR}(H_1^{++} \rightarrow W^+ W^+)$&$9.3 \times 10^{-15}$&$2.7
  \times 10^{-14}$&$5.3 \times 10^{-11}$&$5.1 \times
  10^{-11}$\\ \cline{2-6} 
& $\mbox{BR}(H_1^{++} \rightarrow \ell^+_i \ell^+_j)$&$6.4 \times 10^{-11}$&$1.7
  \times 10^{-12}$&$7.6 \times 10^{-13}$&$2.3 \times
  10^{-9}$\\ \cline{2-6} 
Scenario 3 & $\mbox{BR}(H_2^{++} \rightarrow W^+ W^+)$&$0.03
$&$0.04$&$0.89$&$0.02$\\ \cline{2-6} 
& $\mbox{BR}(H_2^{++} \rightarrow \ell^+_i \ell^+_j)$
&$0.97$&$0.96$&$0.11$&$0.98$\\ \cline{2-6} 
& $\sigma(pp \rightarrow H_1^{++} H_1^{--})$&$\ 0.81$\,fb&
$\ 0.75$\,fb&$\ 0.72$\,fb&$\ 0.83$\,fb\\ \cline{2-6} 
& $\sigma(pp \rightarrow H_2^{++} H_2^{--})$&$\ 3.28$\,fb&
$\ 3.20$\,fb&$\ 2.50$\,fb&$\ 2.54$\,fb\\ 
  \hline
\end{tabular}
\caption{Decay branching ratios and production cross sections for
  doubly-charged scalars for phase $\alpha = 60^{\circ}$. \label{branching3}} 
\end{table}
\clearpage

\newpage
\section{Summary and conclusions}
\label{conc}
We have considered a one-doublet, two-triplet Higgs scenario, with one CP-violating phase in the potential. It is noticed that a larger phase leads to bigger mass-separations between the two doubly-charged mass eigenstates, and also between the states $H_1^{++}$ and $H_2^+$. Consequently, this scenario admits a larger region of the parameter space, when the decay $H_1^{++} \rightarrow H_2^{++} h$ opens up. When it is allowed, this decay often overrides $H_1^{++} \rightarrow H_2^+  W^+$. While the role of the latter decay as a characteristic signal of such models was discussed in our earlier work, we emphasize here that the former mode leads to another interesting signal, arising from $H^{++}_1 \rightarrow \ell^+_i \ell^+_j h$. This would mean that the production of SM-like Higgs together with same-sign dileptons peaking at the mass of the lighter doubly-charged scalar. Such a signal, too, may give us a distinctive signature of a two-triplet scenario at the LHC.

\paragraph{Acknowledgements:}
This work has been
partially supported by the Department of Atomic Energy, 
Government of India, through funding available for
the Regional Centre for Accelerator-Based Particle Physics, 
Harish-Chandra Research Institute. 
We thank Subhadeep Mondal and Nishita Desai for helpful discussions.

\section*{Appendix}
The various elements of $\mathcal{M}_{neut}$, the neutral scalar mass matrix, are as follows : 
\begin{eqnarray}
m_{11} = 2 (b_{22} + \frac{1}{2} (e_{22} - h_{22}) v^2 + 2 (3 d_{22} w_2 ^2 + d_{12} \lvert w_1\rvert ^2 cos2\alpha + 2 (g + g') \lvert w_1\rvert ^2)))
\end{eqnarray}
\begin{eqnarray}
m_{12} = m_{21} = 2 b_{12} + (e_{12} - h_{12}) v^2 + 4 (d_{12} + 2 (g + g')) \lvert w_1\rvert cos\alpha w_2, 
\end{eqnarray}
\begin{eqnarray}
m_{13} = m_{31} = \sqrt{2} v (2 t_2 + (e_{12} - h_{12})\lvert w_1\rvert cos\alpha + (e_{22} - h_{22})w_2), 
\end{eqnarray}
\begin{eqnarray}
m_{14} = m_{41} = 2 d_{12} \lvert w_1\rvert ^2 sin2\alpha ,  
\end{eqnarray}
\begin{eqnarray}
m_{15} = m_{51} = -4 (d_{12} - 2 (g + g')) w_2 \lvert w_1\rvert sin\alpha , 
\end{eqnarray}
\begin{eqnarray}
m_{16} = m_{61} = 0,
\end{eqnarray}
\begin{eqnarray}
m_{22} = 2 (b_{11} + \frac{1}{2} (e_{11} - h_{11}) v^2 + (d_{12} + 2 (g + g')) w_2 ^2 + d_{11} \lvert w_1\rvert ^2 (2 cos^2\alpha + 1)), 
\end{eqnarray}
\begin{eqnarray}
m_{23} = m_{32} = \sqrt{2} v (2 \lvert t_1\rvert cos\beta + (e_{11} - h_{11})\lvert w_1\rvert cos\alpha + (e_{12} - h_{12}) w_2), 
\end{eqnarray}
\begin{eqnarray}
m_{24} = m_{42} = 4 d_{12} w_2 \lvert w_1\rvert sin\alpha , 
\end{eqnarray}
\begin{eqnarray}
m_{25} = m_{52} = 2 d_{11} \lvert w_1\rvert ^2 sin2\alpha , 
\end{eqnarray}
\begin{eqnarray}
m_{26} = m_{62} = 2 \sqrt{2} v \lvert t_1\rvert sin\beta , 
\end{eqnarray}
\begin{eqnarray}
m_{33} =  \frac{1}{4} (2 a + 6 c v^2 + 4 \lvert t_1\rvert \lvert w_1\rvert cos(\alpha + \beta) - h_{11} w_1 ^2 + 4 t_2 w_2 + \\ \nonumber
 2 (e_{12} - h_{12})\lvert w_1\rvert cos\alpha w_2 + (e_{22} - h_{22}) w_2 ^2 - h_{11} \lvert w_1\rvert ^2 sin^2\alpha  + e_{11} \lvert w_1\rvert ^2), 
\end{eqnarray}
\begin{eqnarray}
m_{34} = m_{43} = \sqrt{2} (e_{12} - h_{12}) v \lvert w_1\rvert sin\alpha , 
\end{eqnarray}
\begin{eqnarray}
m_{35} = m_{53} = \sqrt{2} v ((e_{11} - h_{11}) \lvert w_1\rvert sin\alpha - 2 \lvert t_1\rvert sin\beta), 
\end{eqnarray}
\begin{eqnarray}
m_{36} = m_{63} = 2 \lvert t_1\rvert \lvert w_1\rvert sin(\alpha + \beta), 
\end{eqnarray}
\begin{eqnarray}
m_{44} = 2 (b_{22} + \frac{1}{2} (e_{22} - h_{22}) v^2 + d_{22} w_2 ^2 + d_{12} \lvert w_1\rvert ^2 cos2\alpha + 2 (g + g') \lvert w_1\rvert ^2), 
\end{eqnarray}
\begin{eqnarray}
m_{45} = m_{54} = 2 b_{12} + (e_{12} - h_{12}) v^2 + 4 d_{12} \lvert w_1\rvert cos\alpha w_2 , 
\end{eqnarray}
\begin{eqnarray}
m_{46} = m_{64} = 4 \sqrt{2} t_2 v , 
\end{eqnarray}
\begin{eqnarray}
m_{55} = 2 (b_{11} + \frac{1}{2} (e_{11} - h_{11}) v^2 - 2 (d_{12} - 2 (g + g')) w_2 ^2 + 2 d_{11} \lvert w_1\rvert ^2 (2 sin^2\alpha + 1)) , 
\end{eqnarray}
\begin{eqnarray}
m_{56} = m_{65} = 2 \sqrt{2} \lvert t_1\rvert v cos\beta  , 
\end{eqnarray}
\begin{eqnarray}
m_{66} = \frac{1}{4} (2 a + 2 c v^2 - 4 \lvert t_1\rvert \lvert w_1\rvert cos(\alpha + \beta) - (e_{11} - h_{11}) \lvert w_1\rvert ^2 -  4 t_2 w_2 + \\ \nonumber
 2 (e_{12} - h_{12})\lvert w_1\rvert w_2 cos\alpha + (e_{22} - h_{22}) w_2 ^2),
\end{eqnarray}

\end{document}